\documentclass[twocolumn]{aastex62}

\usepackage{times}
\usepackage{graphicx}
\usepackage{amssymb}
\usepackage{amsmath}
\usepackage{pifont}
\usepackage{graphics}
\usepackage{epsfig}
\usepackage{multirow}

\usepackage[normalem]{ulem}

\newcommand{\degrees}{\ensuremath{^{\circ}}}
\newcommand{\bjdtdb}{\ensuremath{\rm {BJD_{TDB}}}}

\newcommand{\teff}{\ensuremath{T_{\rm eff}}}

\newcommand{\loggstar}{\ensuremath{\log{g_\star}}}

\newcommand{\vsini}{\ensuremath{v\sin{I_*}}}

\begin{document}

\title{The KELT Follow-Up Network and Transit False Positive Catalog:\\Pre-vetted False Positives for TESS}

\author[0000-0001-6588-9574]{Karen A.\ Collins}
\affiliation{Harvard-Smithsonian Center for Astrophysics, 60 Garden St, Cambridge, MA, 02138, USA}

\author{Kevin I.\ Collins}
\affiliation{Department of Physics and Astronomy, Vanderbilt University, Nashville, TN 37235, USA}

\author[0000-0002-3827-8417]{Joshua Pepper}
\affiliation{Department of Physics, Lehigh University, 16 Memorial Drive East, Bethlehem, PA 18015, USA}

\author[0000-0002-2919-6786]{Jonathan Labadie-Bartz}	
\affiliation{Department of Physics, Lehigh University, 16 Memorial Drive East, Bethlehem, PA 18015, USA}
\affiliation{Department of Physics and Astronomy, University of Delaware, Newark, DE 19716, USA}

\author[0000-0002-3481-9052]{Keivan G.\ Stassun}
\affiliation{Department of Physics and Astronomy, Vanderbilt University, Nashville, TN 37235, USA}
\affiliation{Department of Physics, Fisk University, 1000 17th Avenue North, Nashville, TN 37208, USA}

\author[0000-0003-0395-9869]{B.\ Scott Gaudi}
\affiliation{Department of Astronomy, The Ohio State University, 140 West 18th Avenue, Columbus, OH 43210, USA}	
\author[0000-0001-6023-1335]{Daniel Bayliss}
\affiliation{Department of Physics, University of Warwick, Gibbet Hill Rd., Coventry, CV4 7AL, UK}
\affiliation{Research School of Astronomy and Astrophysics, Mount Stromlo Observatory, Australian National University, Cotter Road, Weston, ACT, 2611, Australia}

\author[0000-0002-9832-9271]{Joao Bento}
\affiliation{Research School of Astronomy and Astrophysics, Mount Stromlo Observatory, Australian National University, Cotter Road, Weston, ACT, 2611, Australia}

\author[0000-0001-8020-7121]{Knicole D.\ Col\'on}
\affiliation{NASA Goddard Space Flight Center, Exoplanets and Stellar Astrophysics Laboratory (Code 667), Greenbelt, MD 20771, USA}

\author[0000-0002-2457-7889]{Dax  Feliz}
\affiliation{Department of Physics, Fisk University, 1000 17th Avenue North, Nashville, TN 37208, USA}
\affiliation{Department of Physics and Astronomy, Vanderbilt University, Nashville, TN 37235, USA}

% ...

\shorttitle{KELT-FUN and KELT False Positives}

\begin{abstract}
The Kilodegree Extremely Little Telescope (KELT) project has been conducting a photometric survey for transiting planets orbiting bright stars for over ten years.  The KELT images have a pixel scale of $\sim 23\arcsec$ pixel$^{-1}$---very similar to that of NASA's Transiting Exoplanet Survey Satellite (TESS)---as well as a large point spread function, and the KELT reduction pipeline uses a weighted photometric aperture with radius $3\arcmin$.  At this angular scale, multiple stars are typically blended in the photometric apertures.  In order to identify false positives and confirm transiting exoplanets, we have assembled a follow-up network (KELT-FUN) to conduct imaging with higher spatial resolution, cadence, and photometric precision than the KELT telescopes, as well as spectroscopic observations of the candidate host stars.  The KELT-FUN team has followed-up over 1,600 planet candidates since 2011, resulting in more than 20 planet discoveries. Excluding $\sim$450 false alarms of non-astrophysical origin (i.e., instrumental noise or systematics), we present an all-sky catalog of the 1,128 bright stars $(6<V<13)$ that show transit-like features in the KELT light curves, but which were subsequently determined to be astrophysical false positives (FPs) after photometric and/or spectroscopic follow-up observations. The KELT-FUN team continues to pursue KELT and other planet candidates and will eventually follow up certain classes of TESS candidates. The KELT FP catalog will help minimize the duplication of follow-up observations by current and future transit surveys such as TESS.
\end{abstract}


\section{Introduction}

Wide-field surveys for transiting planets are notoriously plagued by astrophysical false positives (FPs). These are due to configurations of stars and/or intrinsic stellar variability that mimic the signal of a transiting planet, i.e. a shallow ($\lesssim 5\%$) dip in apparent brightness of what appears to be a single, isolated star in the survey data, which repeats periodically and has approximately the shape and duration expected for a transiting planet of the observed period. Classification of various types of FPs has been addressed by a number of papers, most notably in \citet{Brown:2003} and \citet{Charbonneau:2004}, but also in \citet{Torres:2004}, \citet{ODonovan:2006}, \citet{Latham:2009}, \citet{Evans:2010} and \citet{Sullivan:2015}. One of the most common astrophysical configurations that can lead to an FP is an eclipsing binary (EB) star system blended in the wide-field survey images with one or more additional (typically brighter) stars.  All else being equal, the contamination of flux from nearby stars and blending of nearby eclipsing binaries (NEBs) in the photometric aperture becomes progressively worse the larger the pixel scale. Also, ground-based surveys have limited photometric precision and thus may not be sensitive enough to detect very shallow secondary eclipses of hierarchical eclipsing systems or blended EBs \citep{Bayliss:2017}.

Because many transit surveys seek to use wide-field optics, so as to monitor large numbers of stars at once, their angular resolution and subsequent pixel scales are typically larger than the sub-arcsec pixels employed by most optical telescopes, which aim to critically sample typical ground-based seeing of a few arcseconds or less. For small-aperture, ground-based transit surveys, the pixel scales can range from $\sim 3.7\arcsec$ to $\sim 36\arcsec$ pixel$^{-1}$. This includes surveys such as TrES \citep{Alonso:2004}, XO \citep{McCullough:2005}, SuperWASP \citep{Pollacco:2006}, HATNet \citep{Bakos:2007}, KELT \citep{Pepper:2007,Pepper:2012}, QES \citep{Alsubai:2011}, HATSouth \citep{Bakos:2013}, NGTS \citep{Wheatley:2017}, and MASCARA \citep{Talens:2017}.  In order to cover a larger area, even space-based transit searches also employ relatively large pixel scales and have angular resolution considerably worse than the theoretical minimum image size set by diffraction limit, such as CoRoT ($2\farcs3$~pixel$^{-1}$; \citealt{Rouan:1998}) and Kepler ($3\farcs98$~pixel$^{-1}$; \citealt{Borucki:2010}).  In particular, the upcoming Transiting Exoplanet Survey Satellite (TESS) mission, which will monitor nearly the entire sky over a period of two years, has pixels that are $21\arcsec$~pixel$^{-1}$ \citep{Ricker:2014}), which will be comparable to or a few times smaller than the angular scale of the point spread function (PSF). Because of the combination of the relatively large size of the PSF and the resulting large photometric apertures required to sample the PSF, limited photometric precision of the survey telescopes, and several astrophysical scenarios that can be confused with transiting exoplanets (see Section \ref{sec:kfun}), the occurrence rate of astrophysical false positives in wide-field transit surveys is high. Therefore transit candidates must typically be confirmed through extensive photometric and spectroscopic follow-up observations. 

Wide-field transit survey follow-up observations have been used to confirm planets for more than twenty years. As a result, thousands of FPs have been identified along with the hundreds of transiting planets detected from the ground-based surveys and thousands of transiting planets detected from space-based surveys. 
Recently, the EBLM Project began to release the FPs from that survey that turned out to be single-lined spectroscopic binaries with low-mass stellar companions \citep{Triaud:2017}.
However, most of the FP detections of most surveys have not been published or otherwise made public. There are likely many reasons for this, but competition between surveys along with the lack of resources needed to compile and disseminate the information, are likely major underlying issues. The lack of FP information exchange between surveys has necessarily caused much duplication of effort in follow-up observation programs to identify false positives when  multiple surveys are searching the same stars for transiting planets. The upcoming TESS mission will survey nearly the entire sky for transiting exoplanets, including large numbers of stars already surveyed by previous wide-field surveys. Since a large number of FPs have already been identified in the TESS fields by previous surveys, the publication of information describing the identified FPs could significantly reduce the amount of follow-up observations required for TESS or any other current or future transit survey. In this paper we present a comprehensive FP catalog from the KELT transit survey to help minimize duplication of follow-up observations for future transit surveys. Furthermore, we present our follow-up process and demonstrate that photometric follow-up by facilities of all sizes can significantly alleviate pressure on more limited spectroscopic resources for future wide-field transit surveys such as TESS.

\section{The KELT Survey and its Similarity to TESS}

The Kilodegree Extremely Little Telescope (KELT; \citealt{Pepper:2003,Pepper:2007}) is a wide-field photometric transit survey operated by Vanderbilt University, The Ohio State University, and Lehigh University. The survey is designed to find transits of extrasolar planets through high precision (better than $\sim1\%$ RMS) photometry of bright stars of magnitudes $8 < V < 10$. This magnitude range was selected to be fainter than that of comprehensive radial velocity (RV) surveys that had largely been vetted for giant transiting planets, but still brighter than most other transit surveys. The reason for that choice is that transiting planets with brighter host stars can be more precisely characterized with fewer follow-up resources, particularly for studies of exoplanet atmospheres. Performing well beyond the design of the survey, KELT has detected transit-like events that warranted follow-up observations of stars of magnitude $6<V<13$.

KELT consists of two robotic telescopes. KELT-North is located at Winer Observatory in Sonoita, Arizona and KELT-South is located at the South African Astronomical Observatory (SAAO) in Sutherland, South Africa. Having installations in both hemispheres allows KELT to survey a large proportion of the entire sky. Each telescope consists of a Mamiya 645 80mm f/1.9 lens with a 42mm aperture, giving a wide field of view of $26\degrees \times 26\degrees$. The lens is mounted in front of a 9\,$\mu$m $4096\times4096$ pixel Apogee CCD camera, giving a pixel scale of $\sim 23\arcsec$ pixel$^{-1}$, which is very close to the TESS pixel scale of $21\arcsec$ pixel$^{-1}$. The camera and lens are both mounted on a Paramount ME robotic mount. A full description of the telescopes and instrumentation can be found in \citet{Pepper:2003,Pepper:2007}.

\begin{figure*}
\begin{center}
\includegraphics[width=1.0\linewidth,trim=0mm 0mm 0mm 0mm,clip]{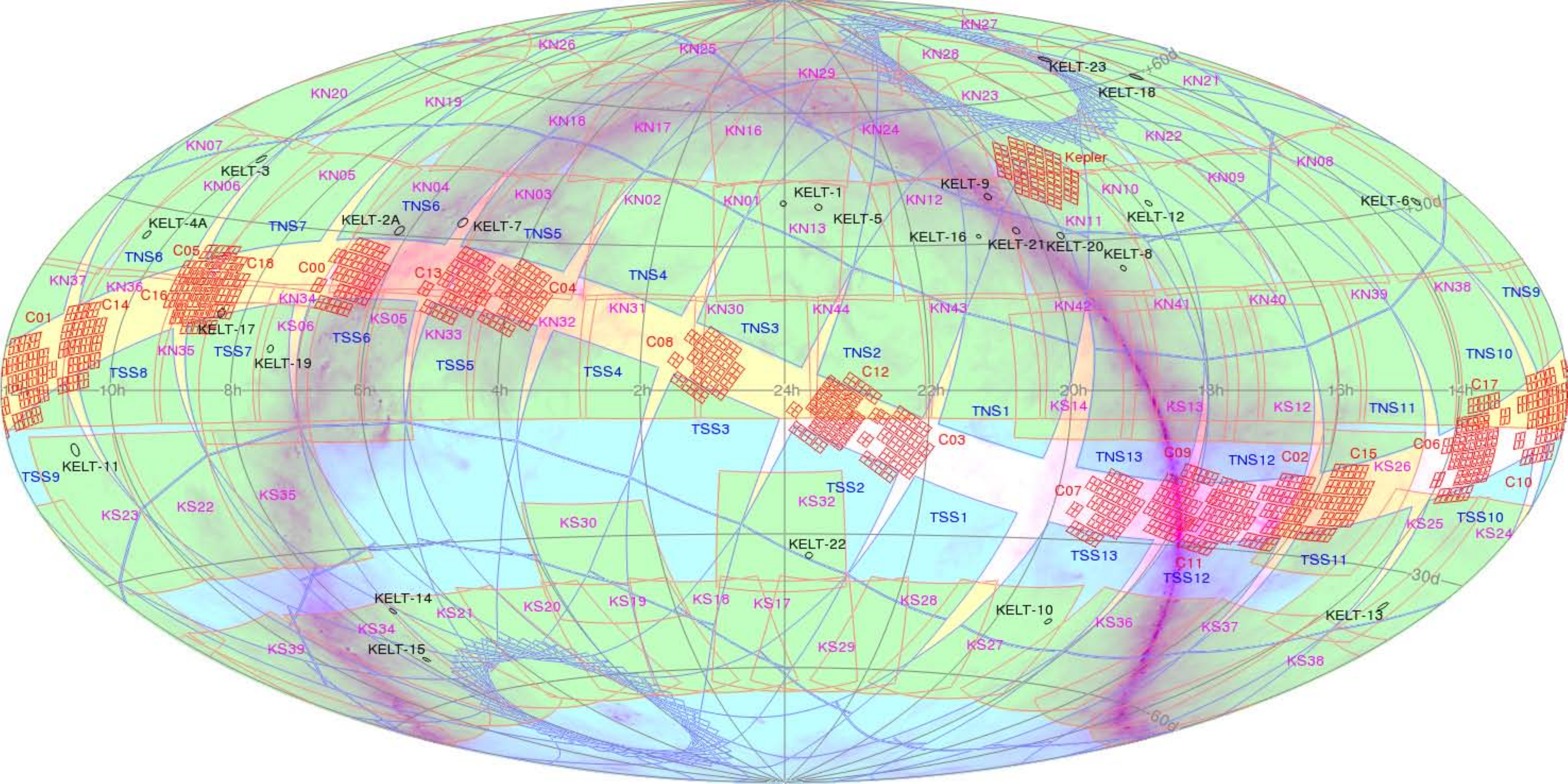}
\caption{The KELT and TESS fields. The KELT fields are outlined in orange and representative TESS fields are outlined in purple. The TESS sector labeled TSS1 represents the actual pointing of TESS Southern Sector 1. The other Sectors are placed relative to Sector 1 and the positions are subject to change. Regions where KELT fields overlap TESS fields are green. Regions where TESS fields have no KELT field overlap are light blue. Regions with no TESS coverage are yellow. The KELT-North fields are labeled KN01 - KN44 and the KELT-South fields are labeled KS05 - KS39. Also shown are the Kepler and K2 fields outlined in red, a model of the galactic plane in magenta, and the locations of the planets discovered by KELT identified by black labels. Fields KN01 - KN13 have been observed for more than ten years as of the publication of this work. Most of the other KELT fields have been observed for three to six years. Figure created with the Montage image mosaic engine \citep{Berriman:2017}.
\label{fig:kelt_fields}}
\end{center}
\end{figure*}

The KELT telescopes have now surveyed more than 70\% of the sky and have discovered transiting planets with transit depths as shallow as $\sim 0.25$\% \citep{Pepper:2017}. Figure \ref{fig:kelt_fields} shows the location of all defined KELT fields (outlined with orange lines) and 26 representative TESS sectors starting near the ecliptic and overlapping at the ecliptic poles (outlined with purple lines). The TESS sector labeled TSS1 represents the actual pointing of TESS Southern Sector 1. The other Sectors are placed relative to Sector 1 and the positions are subject to change. The regions where KELT fields overlap representative TESS sectors appear as green. The regions of TESS fields with no KELT overlap appear as light blue. 

The PSFs of the KELT cameras result in substantial blending of targets with neighboring stars, including blended nearby eclipsing binaries which can masquerade as transiting exoplanets in the KELT photometry as noted above. Also, the limited KELT photometric precision and the various astrophysical FP scenarios described in Section \ref{sec:kfun} cause confusion between transiting exoplanets and false positives. In order to identify false positives and distinguish them from bona fide transiting exoplanets, the KELT project collaborates with a large network of photometric and spectroscopic follow-up observers. The photometric observations are conducted with telescopes and imaging cameras that provide higher spatial resolution, cadence, and photometric precision than the KELT telescopes. The spectroscopic observations provide candidate host star spectroscopic parameters and radial velocity measurements of varying precision. Since ground-based photometry has relative photometric precision limited to of order a millimagnitude, putative secondary eclipses occurring in hierarchical eclipsing systems and blended EBs may not be detected in either the KELT light curves or follow-up light curves \citep{Bayliss:2017}. We describe spectroscopic and photometric techniques to identify these FPs in Section \ref{sec:kfun}.

\subsection{Transit Identification}

The KELT survey identifies, pursues, and validates transiting exoplanet candidates in stages. First, light curves produced by the survey are searched for transit signals and are then subjected to various statistical cuts. All candidates that pass these automated cuts are then manually vetted by the KELT Science Team. This manually-selected subset is then pursued with follow-up observations. It is through these follow-up observations that we identify and categorize the false positives that are the subject of this paper.

\subsubsection{Automated Detection of Transit Candidates}

The KELT-North and KELT-South data reduction pipelines and the process of identification of transit candidates are described in \citet{Siverd:2012} and \citet{Kuhn:2016}, respectively. A short summary is provided here. KELT uses an image subtraction pipeline based on the ISIS software \citep{Alard:1998,Alard:2000}, but with extensive modifications. KELT fields are reduced once every one to three years as new data are acquired, and light curves are produced for all sources identified in the images. The source list is then cross-matched against the Tycho \citep{Hog:2000} and UCAC4 \citep{Zacharias:2013} catalogs. We then use a reduced proper motion (RPM) cut \citep{Gould:2003b,Collier:2007} to identify and remove giant stars before conducting the search for transit signals. However, that process is not perfect -- some giant stars have especially large proper motions or are incorrectly measured in proper motion catalogs, and thus some giants thus make it through the RPM cut. A search is then performed on the light curves of all the stars that passed the RPM cut with the Box-fitting Least Squares (BLS) algorithm \citep{Kovacs:2002} to identify targets exhibiting transit-like signals. The BLS algorithm provides several signal detection metrics that are used to perform automated cuts when assembling the initial list of transiting planet event candidates for each reduced field. The metrics (see \citealt{Hartman:2016} for detailed definitions) and typical limits are specified in Table \ref{tbl:BLS_Selection_Criteria}.

\begin{table}[hb]
\centering
\footnotesize
\caption{Typical KELT BLS selection criteria}
\label{tbl:BLS_Selection_Criteria}
{\setlength{\tabcolsep}{0.40em}
\begin{tabular}{lllll}
    \hline
    BLS Statistic & Selection \\
    (see \citealt{Hartman:2016})              & Criteria  \\
    \hline
    Signal detection efficiency\dotfill & SDE $>$ 7.0\\
    Signal to pink-noise\dotfill        & SPN $>$ 7.0\\
    Transit depth\dotfill               & $\delta <$ 0.05\\
    $\chi^2$ ratio\dotfill & $\frac{\Delta\chi^2_{\rm transit}}{\Delta\chi^2_{\rm inverse~transit}} > 1.5$\\
    Duty cycle\dotfill                  & q $<$ 0.1\\
    $|\log \frac{\rho_{\rm obs}}{\rho_{\rm calc}}|$\dotfill    &   $\leq$ 1.0\\ 
    Fraction from one night\dotfill     & $f_{1n}<0.8$\\
    \hline
    
\end{tabular}}
\end{table}

\subsubsection{Human Vetting of Transit Candidates} \label{sec:hum_vet}

\begin{figure*}[t]
\begin{center}
\includegraphics[width=1.0\linewidth,angle=0,trim=0mm 0mm 0mm 0mm,clip]{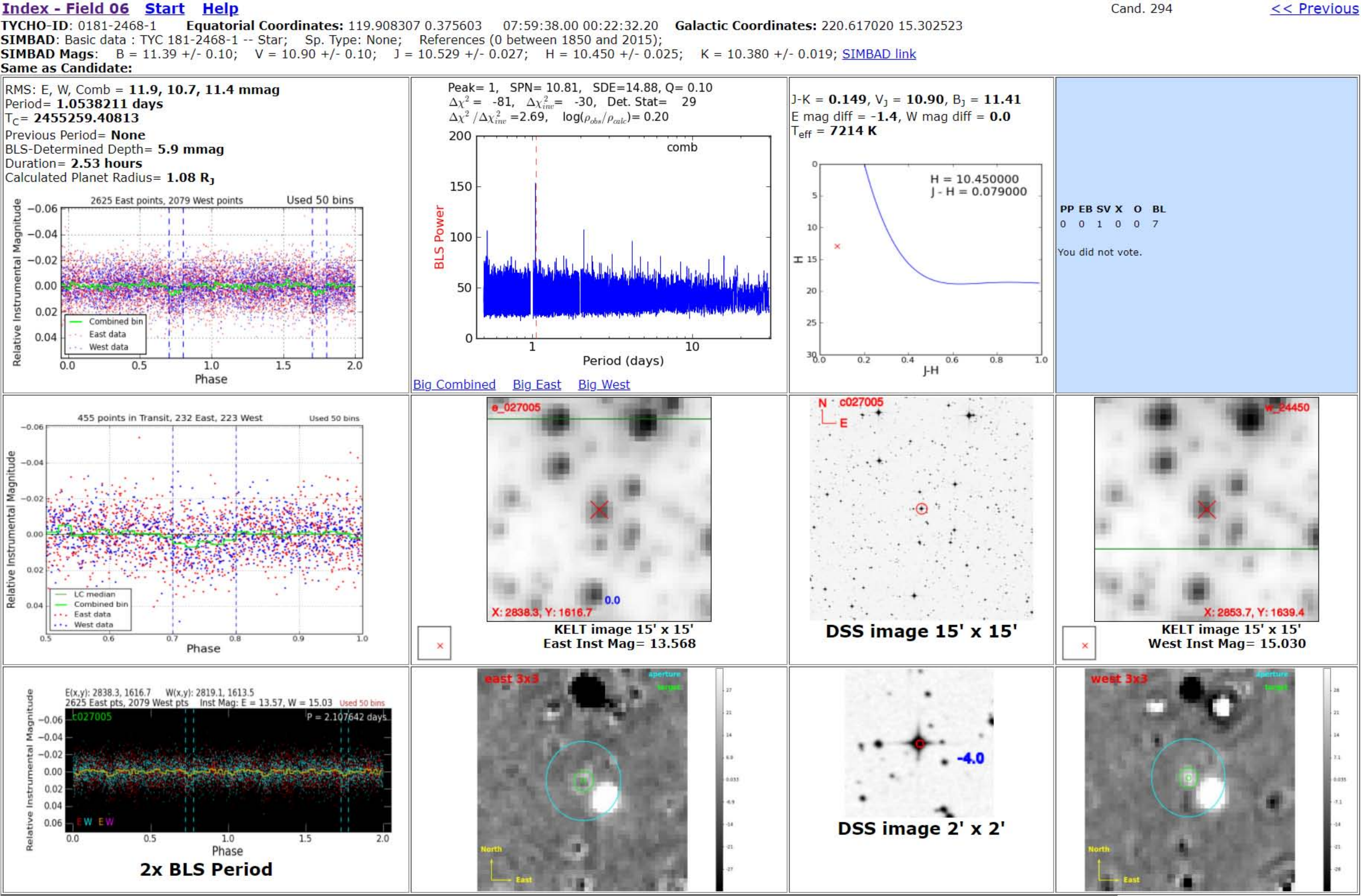}
\caption{Example of a portion of a KELT candidate page. Candidate pages are created for all KELT detected transit-like events that pass the automated statistical cuts. The light curve plots, BLS periodogram, field images, image variability centroids, and various catalog data provide the information needed for human vetting to determine the likelihood of a signal being caused by a genuine exoplanet transiting the star, rather than by an FP. The second and forth panels from the left in the bottom row show difference images of the in-transit versus out-of-transit KELT images. Pixels whose variability correlate with the light curve transit times show strong signals in white. Since the variability is located significantly off-center from the target star in this example, the source of the transit signal is in fact a nearby eclipsing binary.}
\label{fig:candidate_page}
\end{center}
\end{figure*}

Following the automated candidate selection process, the KELT Science Team examines the candidates in further detail, in order to select a subset of targets to be pursued with follow-up observations. This process begins with the creation of an online candidate web page for each object that passes the automated statistical cuts. An example of a portion of a candidate page is shown in Figure \ref{fig:candidate_page}. The candidate page is designed to give KELT Science Team members an overall impression of the likelihood of the transit detection being astrophysically real (as opposed to a spurious signal caused by noise or telescope systematics), and, if real, the likelihood of the signal being caused by a genuine exoplanet transiting the star, rather than by an EB or some other type of FP. Various plots, statistics, and other information (\textit{e.g.} from the SIMBAD Astronomical Database\footnote{http://simbad.u-strasbg.fr/simbad/}, sky images with high spatial resolution, and measurements from catalogs) assist in this endeavor. Each KELT Science Team member inspects each candidate page for all objects that pass the automated selection criteria in a given field, and based on the member's best interpretation of the data, votes if they are in favor of pursuing the candidate with follow-up observations. Team members can also add comments to each target to explain the reasoning for their choice, or raise questions or concerns about the candidate. 

After the voting phase, a group vetting conference call is held. The purpose of this conference call is to discuss the merits of, or problems with, each candidate for which over half of the KELT Science Team has voted in favor of pursuing. At this stage, the KELT Science Team decides whether or not to request follow-up data for each target. For each target being pursued with follow-up observations, a priority is assigned. Higher priorities are given to candidates that are scientifically valuable (\textit{e.g.} bright host stars), have a high likelihood of being genuine transiting exoplanets, and/or have long orbital periods (since transit events are relatively rare for longer periods, it is desirable to observe these when the opportunity arises). The default follow-up observing strategy begins with requesting time-series photometric observations of the transit with a seeing-limited telescope. However, some candidates are also pursued immediately with spectroscopic follow-up observations. For example, bright, long-period, isolated candidates may warrant immediate spectroscopic follow-up. This is because an EB is a likely FP scenario for giant planets, and spectroscopic observations are typically more efficient at confirming or ruling out an EB hypothesis for long period candidates compared to photometric observations, since typically only two RV measurements are required to determine that the orbiting companion has a stellar mass, and the precise timing of these measurements is not essential, whereas it may take many months for the predicted transit of a long period candidate to be observable using photometric follow-up resources. 

The end product is a list of candidates, each with their own priority, follow-up observing strategy, and notes to the observers. These candidates are generally made available to follow-up observers the following day, and the next phase of our candidate vetting process begins.
\\

\section{The KELT Follow-Up Network}\label{sec:kfun}

\subsection{KELT-FUN Members and Followup Framework}

\begin{figure*}
\begin{center}
\includegraphics[width=1.0\linewidth,trim=0mm 0mm 0mm 0mm,clip]{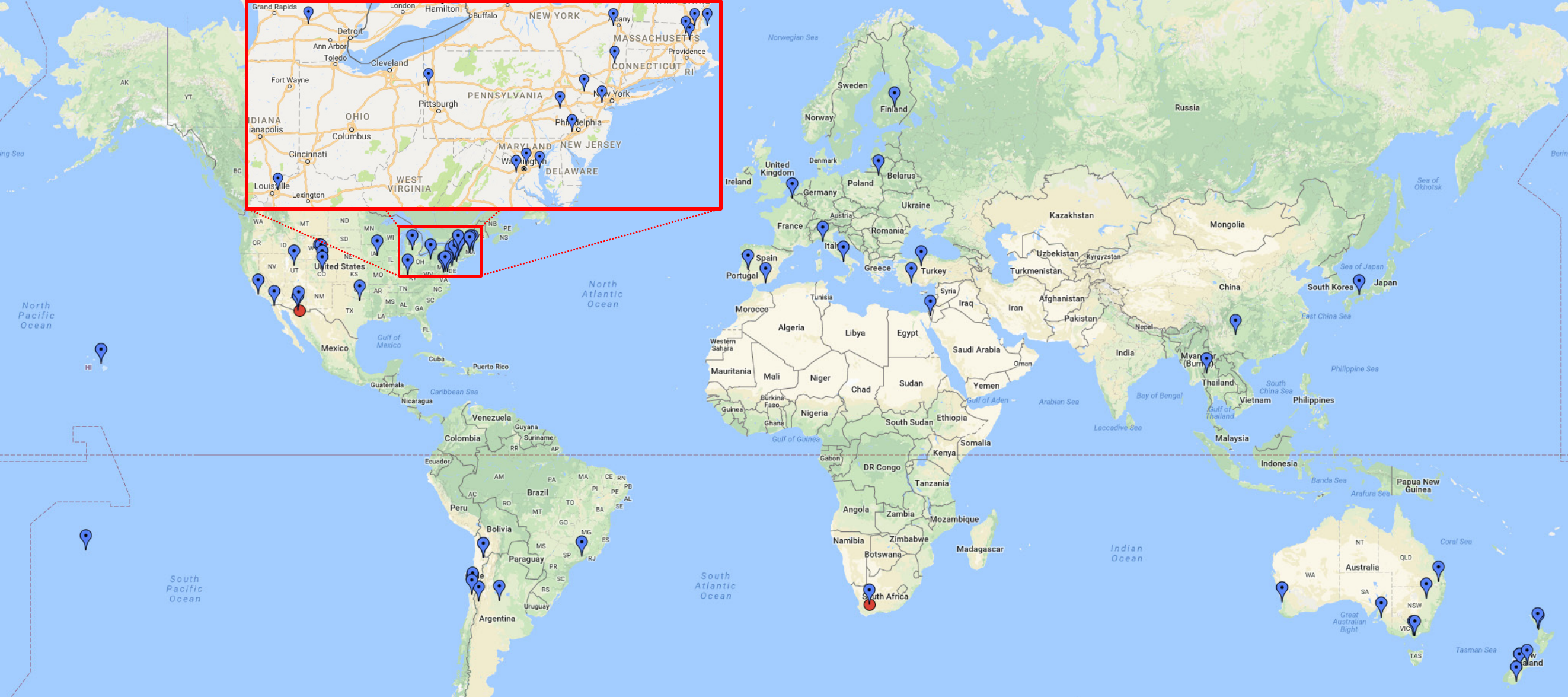}
\caption{The KELT-FUN observatory locations. The map illustrates the longitudinal and latitudinal coverage of the KELT-FUN observatories (blue location markers) and the two KELT telescopes (red dots). The inset shows an expanded view of the eastern United States observatories. Map data: Google. 
\label{fig:kelt-fun_map}}
\end{center}
\end{figure*}

The primary goal of the KELT Follow-up Network (KELT-FUN) is to confirm and characterize transiting exoplanets orbiting bright stars, but additional science projects investigating eclipsing binaries and other variable stars are also pursued. The members of KELT-FUN are a mix of professional, student, and highly capable citizen astronomers distributed across the globe. Figure \ref{fig:kelt-fun_map} shows the longitudinal and latitudinal distribution of the KELT-FUN observatories. Table \ref{tab:kfun} lists the KELT-FUN participating observatories and instrumentation specifications. KELT-FUN started operations in the spring of 2011 when the first KELT transiting planet candidates were extracted from the KELT data and vetted by the KELT Science Team. 

\begin{figure}
\begin{center}
\includegraphics[angle=90,width=1.0\linewidth,trim=0mm 0mm 0mm 0mm,clip]{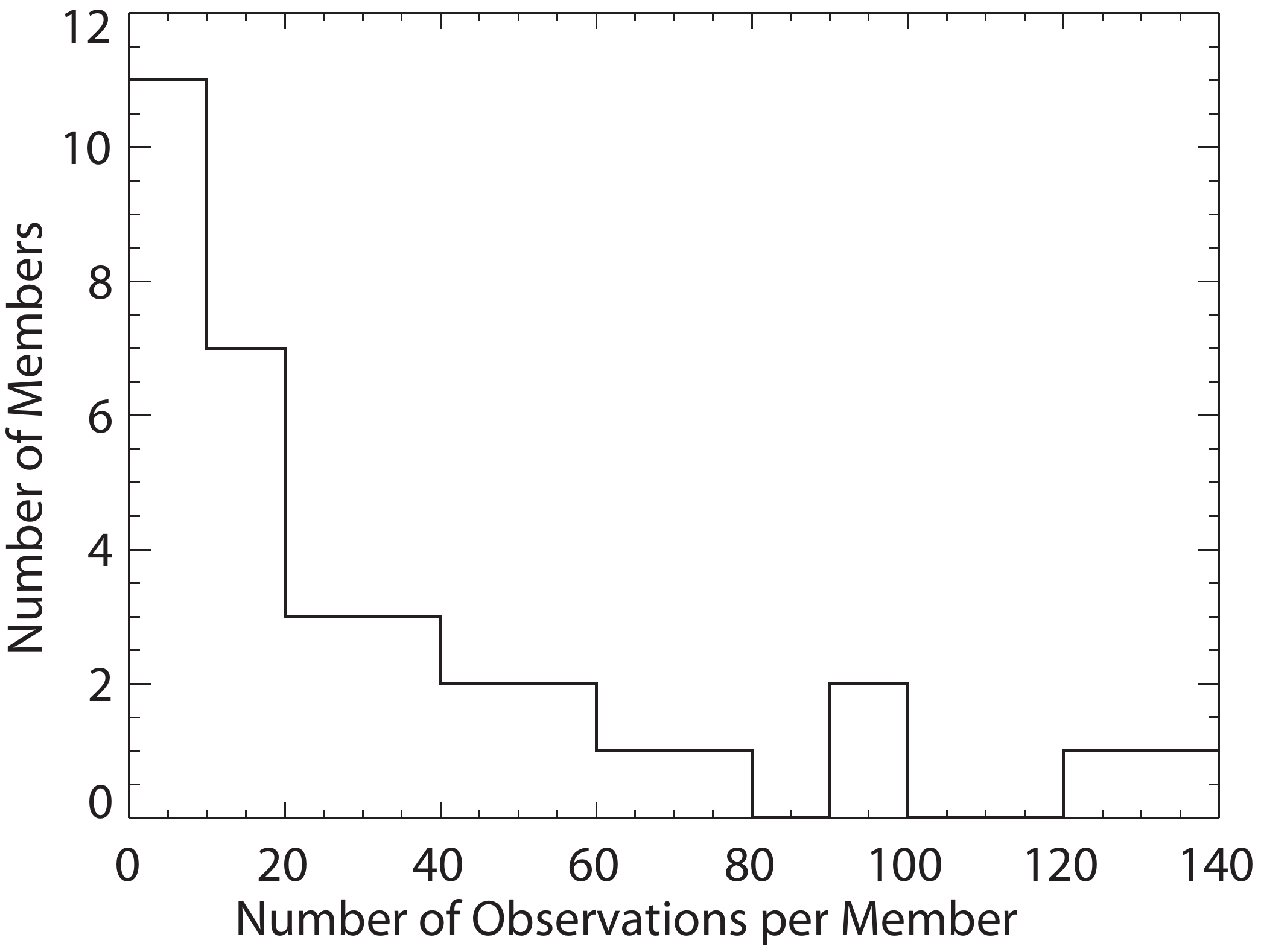}
\caption{Distribution of the number of photometric followup observations per KELT-FUN member, for the calendar year 2016. Note that we include here only KELT-FUN members who contributed at least one photometric followup observation in 2016; the first bin represents members who contributed between 1 and 9 observations.
\label{fig:kfun_hist}}
\end{center}
\end{figure}

Figure~\ref{fig:kfun_hist} displays the relative distribution of observations by 
members of KELT-FUN during the calendar year 2016.  In this case, each 
observation represents a single night of time-series photometry taken by 
one of the collaborating KELT-FUN members.  There are 1018 observations 
represented here by 34 separate members.  Note that participation by 
members can wax and wane over time due to weather, equipment problems, 
or time availability of observers.  Additional members joined the 
collaboration since the end of 2016, and some members who were highly 
active in previous years have since had low productivity.
We see a roughly power-law distribution in which a handful of 
institutions provide the bulk of the observations.  Nevertheless, 20\% of 
the observations were provided by 21 of the collaborating institutions 
who observed fewer than 25 nights in the year.  As time progresses, more 
members become increasingly skilled and efficient at observing.
The 1018 observations included in Figure~\ref{fig:kfun_hist} do not each represent a 
confirmed planet or expired candidate.  Frequently, multiple nights of  
observation are needed to identify certain types of False Positives, or 
to determine that a candidate was a False Alarm.  Such determinations 
are a collective use of multiple observations, and usually cannot be 
ascribed to a single observer.

As previously described, the KELT telescope pixels are large ($\sim23\arcsec$) and the KELT photometric aperture has an effective radius of $3\arcmin$ (although see \citealt{Siverd:2012} for details on how the KELT pipeline uses a weighted aperture). The size of the effective aperture makes it likely that multiple stars are blended with a typical KELT target star. Furthermore, the KELT light curves have limited photometric precision, and some astrophysical FP configurations can not be identified with photometric observations alone. In general, the FP transit-like detections in the KELT light curves can be caused by several scenarios including:
\begin{itemize}
\item A candidate star that is an eclipsing binary (EB), in which the large depth of the primary eclipse of the EB is diluted to the depth of a possible transiting planet through blending with multiple other sources in the KELT photometric aperture. KELT-FUN telescopes with higher spatial resolution are able to distinguish the multiple sources and identify the large depth of the EB as being non-planetary.
\item A pulsating or rotational variable star that appears to be a transit candidate when observed at lower precision, typically due to BLS picking up an alias of the true variable signal.
\item A non-varying candidate star that is blended with a nearby eclipsing binary (NEB) which is inside or close to the KELT aperture. KELT-FUN telescopes with higher spatial resolution are able to distinguish the multiple sources and identify the other source as the origin of the apparent transit signal.
\item A fully blended eclipsing binary (BEB). These are cases similar to the previous one, with the difference being that the candidate star has a angular separation that is so close to the other sources that even the follow-up observations are unable to spatially distinguish the separate sources. In these cases, the stellar nature of the eclipses can be detected if the eclipse depth varies in different bandpasses because the color of the target star and EB are sufficiently different so that the fractional contamination varies with wavelength. This category includes chance alignments and hierarchical stellar systems, since we do not attempt to differentiate between bound and unbound systems.
\item An eclipsing binary system in which the secondary star is small enough in comparison to the primary to produce a primary eclipse with a depth consistent with a planetary transit, even without dilution by blending with other stars. This can be caused by a configuration of a giant primary and a main-sequence secondary, or a more massive main sequence primary star and a lower-mass secondary main sequence star.
\item A grazing eclipsing binary system, in which the depth of the primary eclipse is small enough to be consistent with a transiting planet. Grazing eclipses generally have a V-shaped morphology, but with the typical photometric precision of KELT, may require additional follow-up to differentiate the trapezoidal morphology typically expected from a transiting planet (or in general an eclipsing companion with radius much smaller than the primary), from the V-shaped morphology expected from a grazing system.  We note that grazing planetary systems can also have a V-shaped morphology, e.g., \citet{Odonovan:2007}.  
\item Finally, there are transit candidate false alarms (FAs) caused by instrumental or systematic noise, and are therefore non-astrophysical FPs. We do not discuss these further, except to note that with nearly 30\% of all our FPs being non-astrophysical, the efficiency of the KELT-FUN network at disposing of these FAs has been a particularly important resource for the KELT survey. 
\end{itemize}

KELT-FUN includes both photometric and spectroscopic follow-up observers. Although the collaboration includes both kinds of observers in communications and analysis efforts, the way that transit candidates are selected for each type of follow-up, and the use of the online tools we have developed are quite different for the different types of observing. The next three sections refer almost exclusively to the operations of photometric follow-up operations.

\begin{table*}[bp]
\centering
\scriptsize
{\setlength{\extrarowheight}{0.7pt}
\begin{tabular}{|l|l|c|c|c|c|c|c|}
\hline
Observatory/Telescope & 
Institution & Latitude & Longitude & Altitude & Aperture & FOV & Scale\vspace{-0.03cm}\\ 
  & & & & (m) & (m) & (arcmin) & $\frac{\rm arcsec}{\rm pixel}$ \\  \hline \vspace{-0.03cm}
AAT/UCLES &  Australian Astronomical Observatory & -31.2770 & 149.0661 & 1100 & 3.9 & spectrograph &  \\  \hline \vspace{-0.03cm}
ANU 2.3m/WiFeS Integral Field Spectrograph & Australian National University & -31.2770 & 149.0661 & 1100 & 2.3 & spectrograph &   \\ \hline \vspace{-0.03cm}
Automated Planet Finder (APF) &  Lick Observatory & 37.3414 & 121.6429 & 4200 & 2.4 & spectrograph & \\ \hline \vspace{-0.03cm}
Euler 1.2m/CORALIE & Geneva Observatory & -29.2594 & -70.7331 & 2400 & 1.2 & spectrograph & \\ \hline \vspace{-0.03cm}
FLWO/TRES & CfA/SAO & 31.6811 & -110.8783 & 1524 & 1.5 & spectrograph &  \\  \hline \vspace{-0.03cm}
Keck I/HIRES & Mauna Kea Observatory & 19.8264 & -155.4742 & 4145 & 10.0 & spectrograph & \\ \hline \vspace{-0.03cm}
Large Binocular Telescope/PEPSI & Mount Graham International Obs. & 32.7013 & -109.8891 & 3221 & $2\times8.4$ & spectrograph & \\ \hline \vspace{-0.03cm}
McDonald/Harlan J. Smith Telescope (HJST) & University of Texas at Austin & 30.6715 & -104.02261 & 2076 & 2.7 & spectrograph  &   \\ \hline  \vspace{-0.03cm}
FLWO/KeplerCam &  CfA/SAO & 31.6811 & -110.8783 & 1524 & 1.2 & 23.1’x23.1’ & 0.37 \\ \hline  \vspace{-0.03cm}
Peter van de Kamp Observatory (PvdKO) & Swarthmore College & 39.9071 & -75.3556 & 65 & 0.6 & 26.1x26.1 & 0.38 \\ \hline  \vspace{-0.03cm}
Moore Observatory RC (MORC) & \multirow{3}{*}{University of Louisville} & 38.3444 & -85.5289 & 229 & 0.6 & 26.6x26.6 & 0.39  \\ \cline{1-1} \cline{3-8}  \vspace{-0.03cm}
Moore Observatory CDK20N &  & 38.3444 & -85.5289 & 229 & 0.5 & 36.9x36.9 & 0.54 \\ \cline{1-1} \cline{3-8}  \vspace{-0.03cm}
Mt Lemmon/UL Manner Telescope (ULMT) &  & 32.4424 & -110.7888 & 2792 & 0.6 & 26.8x26.8 & 0.39 \\ \hline \vspace{-0.03cm}
Mt. Kent/CDK20S & \multirow{2}{*}{U. Louisville/U. Southern Queensland} & -27.7979 & 151.8554 & 682 & 0.5 & 36.9x36.9 & 0.54 \\ \cline{1-1} \cline{3-8} \vspace{-0.03cm}
Mt. Kent/CDK700 &  & -27.7979 & 151.8554 & 682 & 0.7 & 27.3x27.3 & 0.40 \\ \hline \vspace{-0.03cm}
Crow Observatory & Crow Observatory & 39.2 & -7.2 & 460 & 0.3048 & 23x18 & 0.85 \\ \hline \vspace{-0.03cm}
Westminster College Observatory & Westminster College & 41.1176 & -80.3317 & 327 & 0.35 & 24x16 & 0.45 \\ \hline \vspace{-0.03cm}
Kutztown Observatory & Kutztown University & 40.5113 & -75.7858 & 122 & 0.6096 & 13.0x19.5 & 0.72 \\ \hline \vspace{-0.03cm}
Whitin Observatory & Wellesley College & 42.2953 & -71.3067 & 141 & 0.6096 & 20x20 & 0.58 \\ \hline \vspace{-0.03cm}
DEMONEXT - Winer Observatory & Ohio State University & 31.6656 & -110.6018 & 1515.7 & 0.5 & 31x31 & \\ \hline \vspace{-0.03cm}
Shaw Observatory & Shaw Observatory & -31.8944 & 115.9303 & 0 & 0.3556 &  &  \\ \hline \vspace{-0.03cm}
Ellin Bank Observatory &  & -38.2447 & 145.9600 & 0 & 0.3175 & 20.2’x13.5’ & 1.12 \\ \hline \vspace{-0.03cm}
Harlingten San Pedro &  & -22.9167 & -68.2000 & 2400 & 0.5 &  &  \\ \hline \vspace{-0.03cm}
PEST & Perth Exoplanet Survey telescope & -31.9925 & 115.7983 & 19 & 0.3 & 31x21 & 1.20 \\ \hline \vspace{-0.03cm}
ICO & Ivan Curtis Observatory & -34.8845 & 138.6309 & 44 & 0.235 & 19x15 & 0.62 \\ \hline \vspace{-0.03cm}
 \multirow{2}{*}{Red Buttes Observatory} & \multirow{2}{*}{University of Wyoming} & 41.1764 & -105.5740 & 2246 & 0.61 &  25x25 & 0.37 \\ \cline{3-8} \vspace{-0.03cm}
 &  & 41.1764 & -105.5740 & 2246 & 0.61 & 9x9 & 0.53 \\ \hline \vspace{-0.03cm}
MBA Observatory & Montgomery Bell Academy & 35.6772 & -85.6089 & 538 & 0.6096 & 19.9x19.9 & 0.45 \\ \hline \vspace{-0.03cm}
GMU Observatory & George Mason University & 38.8526 & -77.3044 & 95 & 0.8128 & 22.2x22.2 & 0.39 \\ \hline \vspace{-0.03cm}
\multirow{2}{*}{Pratt Observatory} & \multirow{5}{*}{Brigham Young University} & 40.2497 & -111.6489 & 1371 & 0.4064 & 16.6x16.6 & 0.37 \\ \cline{3-8} \vspace{-0.03cm}
 &  & 40.2470 & -111.6503 & 1357 & 0.2 & 25.7x17.3 & 0.72 \\ \cline{1-1} \cline{3-8} \vspace{-0.03cm}
\multirow{2}{*}{West Mountain Observatory} &  & 40.0875 & -111.8256 & 2120 & 0.32 & 17.9 x 12.0 & 0.49 \\ \cline{3-8} \vspace{-0.03cm}
 &  & 40.0875 & -111.8256 & 2120 & 0.91 & 21x21 & 0.61 \\ \hline \vspace{-0.03cm}
Harlingten Observatory - New Mexico &  & 31.9469 & -108.8975 & 1402 & 0.4 &  &  \\ \hline \vspace{-0.03cm}
\multirow{2}{*}{Canis Mayor Observatory} & \multirow{2}{*}{} & 44.1044 & 10.0078 & 0 & 0.254 &  &  \\ \cline{3-8} \vspace{-0.03cm}
 &  & 44.1044 & 10.0078 & 0 & 0.4 &  &  \\ \hline \vspace{-0.03cm}
\multirow{2}{*}{Salerno University Observatory} & \multirow{2}{*}{University of Salerno} & 40.7750 & 14.7889 & 300 & 0.35 & 14.4x10.8 & 0.54 \\ \cline{3-8} \vspace{-0.03cm}
  &  & 40.7750 & 14.7889 & 300 & 0.60 & 20.8x20.8 & 0.61 \\ \hline  \vspace{-0.03cm}
Haleakala Observatory FTN & \multirow{7}{*}{Las Cumbres Observatory (LCO)} & 20.7069 & -156.2572 & 3055 & 2 & 10x10 &  \\ \cline{1-1} \cline{3-8} \vspace{-0.03cm}
BOS &  & 34.6876 & -120.0390 & 500 & 0.8 & 14.7x9.8 & 0.57 \\ \cline{1-1} \cline{3-8} \vspace{-0.03cm}
ELP &  & 30.6700 & -104.0200 & 2070 & 1 & 26.5x26.5 & 0.46 \\ \cline{1-1} \cline{3-8} \vspace{-0.03cm}
LSC 3x &  & -30.1674 & -70.8048 & 2198 & 1 & 26.5x26.5 & 0.39 \\ \cline{1-1} \cline{3-8} \vspace{-0.03cm}
CPT 3x &  & -32.3800 & 20.8100 & 1460 & 1 & 15.8x15.8 & 0.24 \\ \cline{1-1} \cline{3-8} \vspace{-0.03cm}
COJ 2x &  & -31.2733 & 149.0710 & 1116 & 1 & 15.8x15.8 & 0.24 \\ \cline{1-1} \cline{3-8} \vspace{-0.03cm}
FTS &  & -31.2733 & 149.0710 & 1116 & 2 & 10x10 &  \\ \hline \vspace{-0.03cm}
\multirow{2}{*}{Okayama Astrophysical Observatory} & \multirow{2}{*}{National Astronomical Obs. of Japan} & \multirow{2}{*}{34.577} & \multirow{2}{*}{133.594} & \multirow{2}{*}{372} & 1.88 & 6.1x6.1 & 0.36 \\ \cline{6-8} \vspace{-0.03cm}
 &  &  &  &  & 0.50 & 26x26 & 1.50 \\ \hline \vspace{-0.03cm}
Myers T50 &  & -31.2733 & 149.0644 & 1165 & 0.43 & 15.5x15.5 & 0.92 \\ \hline \vspace{-0.03cm}
Mt. John Observatory & University of Canterbury & -43.9856 & 170.4650 & 1029 & 0.61 & 14x14 & 0.55 \\ \hline \vspace{-0.03cm}
Hazelwood Observatory &  & -38.2994 & 146.4239 & 105 & 0.32 & 20.0x13.9 & 1.10 \\ \hline \vspace{-0.03cm}
Adams Observatory & Austin College & 33.6471 & -96.5988 & 254 & 0.61 & 26x26 & 0.38 \\ \hline \vspace{-0.03cm}
\multirow{2}{*}{Ankara University Kreiken Observatory} & \multirow{2}{*}{Ankara University} & 39.8436 & 32.7792 & 1250 & 0.4 & 11x11 &  \\ \cline{3-8} \vspace{-0.03cm}
 &  & 39.8436 & 32.7792 & 1250 & 0.35 & 12x12 & \\ \hline \vspace{-0.03cm}
CU Sommers-Bausch Observatory & University of Colorado - Boulder & 40.0037 & -105.2625 & 1653 & 0.61 & 25x25 &  \\ \hline \vspace{-0.03cm}
OPD Observatory & OPD Observatory & -22.5353 & -45.5828 & 1864 & 0.6 & 10x10 &  \\ \hline \vspace{-0.03cm}
Conti Private Observatory &  & 38.9301 & -76.4883 & 0 & 0.28 & 14.4x11.5 &  \\ \hline \vspace{-0.03cm}
Spot Observatory & Spot Observatory & 35.8847 & -87.5653 & 225 & 0.6096 & 26.8x26.8 &  \\ \hline \vspace{-0.03cm}
CGHome Observatory &  & 43.7928 & 10.4747 & 40 & 0.2 & 59x39 & \\ \hline \vspace{-0.03cm}
AAI - William Miller Sperry Observatory & Union County College, Cranford, NJ & 40.6679 & -74.3201 & 25 & 0.609 & 19.5x13.4 &  \\ \hline \vspace{-0.03cm}
Phillips Academy Observatory & Phillips Academy & 42.6475 & -71.1297 & 100 & 0.4 & 30x30 & 0.89 \\ \hline \vspace{-0.03cm}
Acton Sky Portal &  & 42.4550 & -71.4349 & 60 & 0.355 & 17.5x11.7 &  \\ \hline \vspace{-0.03cm}
Star View Hill Observatory &  & 40.9603 & -74.9461 & 220 & 0.635 & 15x10 & \\ \hline \vspace{-0.03cm}
\multirow{3}{*}{UMD Observatory} & \multirow{3}{*}{University of Maryland} & \multirow{3}{*}{39.0021} & \multirow{3}{*}{-76.956} & \multirow{3}{*}{70} & 0.18 & 32x21.5 &  \\ \cline{6-8}  \vspace{-0.03cm}
 &  &  &  &  & 0.15 & 37.3x25.1 &  \\ \cline{6-8} \vspace{-0.03cm}
 &  &  &  &  & 0.355 & 12.1x8.1 & \\ \hline \vspace{-0.03cm}
Grant O. Gale Observatory & Grinnell College & 41.7556 & -92.7198 & 310 & 0.6096 & 13x13 & 0.37 \\ \hline \vspace{-0.03cm}
SkyNet & University of North Carolina & various & various & various & various & various & various \\ \hline \vspace{-0.03cm}
Rarotonga Observatory & Rarotonga Observatory & -21.2093 & -159.8133 & 32 & 0.25 & 19x19 &  \\ \hline \vspace{-0.03cm}
El Sauce Observatory & El Sauce Observatory & -30.4711 & -70.7650 & 1600 & 0.356 & 18.5’x12.3’ &  \\ \hline \vspace{-0.03cm}
Estaci\'on Astrof\'isica de Bosque Alegre (EABA) & Observatorio Astron\'omico de C\'ordoba & -31.5983 & -64.5467 & 1250 & 1.54 & 17x17 & 0.25 \\ \hline \vspace{-0.03cm}
Pukekohe Observatory &  & -37.1881 & 174.9092 & 41 & 0.3 & 14x12 & 0.30 \\ \hline \vspace{-0.03cm}
iDK & Mt. Stuart Observatory & -46.0227 & 169.8474 & 361 & 0.3175 & 16.2x24 &  \\ \hline \vspace{-0.03cm}
TRT-TNO & Thai National Observatory & 18.5737 & 98.4823 & 2457 & 0.5 & 23.4x23.4 & 0.68 \\ \hline \vspace{-0.03cm}
TRT-GAO & Yunnan Observatory & 26.6955 & 105.031 & 3193 & 0.7 & 20.9x20.9 & 0.61 \\ \hline 
AstroLAB IRIS &  & 50.818 & -2.910 & 39 & 0.175 & 37.3x37.3 & 1.10\\ \hline 
\end{tabular}}
    \caption{KELT Follow-Up Network Telescopes and Instrumentation}
\label{tab:kfun}
\end{table*}

\subsection{Follow-up Photometry}

\subsubsection{Planning Photometric Observations}

\begin{figure*}
\begin{center}
\includegraphics[width=1.0\linewidth,trim=0mm 0.5mm 0mm 0mm,clip]{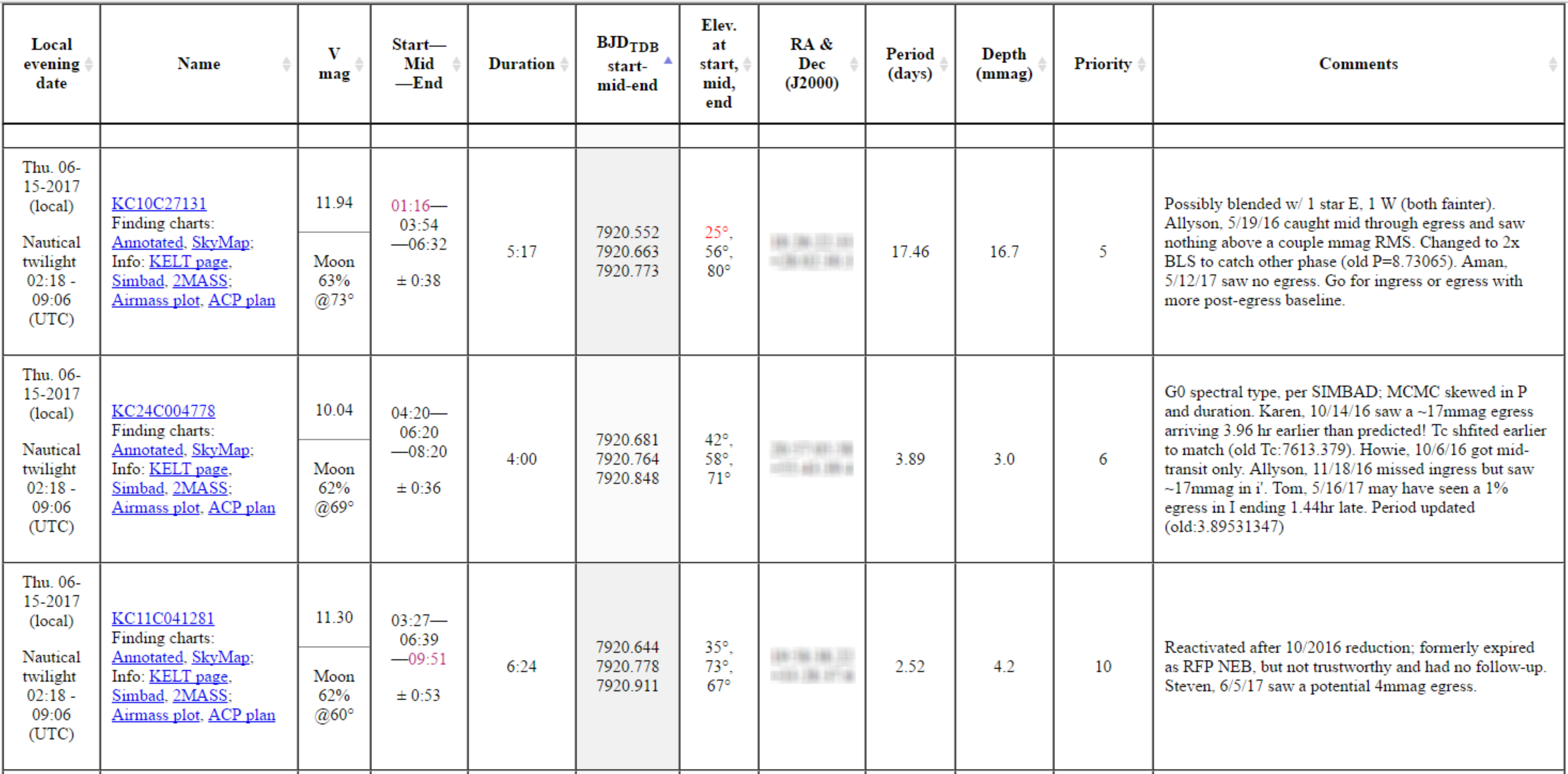}
\caption{The KELT Transit Finder online portal. KELT-FUN team members select events to observe using the KELT Transit Finder tool. The output filter settings include geographical location of the observatory, date range, target elevation above horizon at ingress and/or egress, dark time at the observatory, transit depth, host star brightness, and target priority. Each observable event is described by a row of data in the output which includes the object name and a comprehensive set of observational information for the target.
\label{fig:ktf}}
\end{center}
\end{figure*}

A KELT Transit Finder (KTF) web tool, based on TAPIR \citep{Jensen:2013}, is available to the KELT-FUN team to assist in planning follow-up photometric observations. An observer enters an observatory location, a range of dates to search, and various filtering options to produce a list of observable KELT candidates that will be transiting during a particular observing window at their location. An example of the data display provided by the KTF is shown in Figure \ref{fig:ktf}. Each observable event is described by a row of output data which includes object RA and Dec (J2000), time and elevation range of the event, $V$ magnitude, moon brightness and separation from the target, event period, duration, depth in the KELT aperture, priority ranking, and links to finding charts and other online resources. Observing notes provide a summary of any previous follow-up observations, and suggestions for the next observations. Event times that occur during daylight, elevations that are below a selected threshold, high moon illumination, and close moon proximity are all highlighted with red or magenta text. 

\begin{figure*}
\begin{center}
\includegraphics[width=1.0\linewidth,trim=0mm 0.0mm 0mm 0mm,clip]{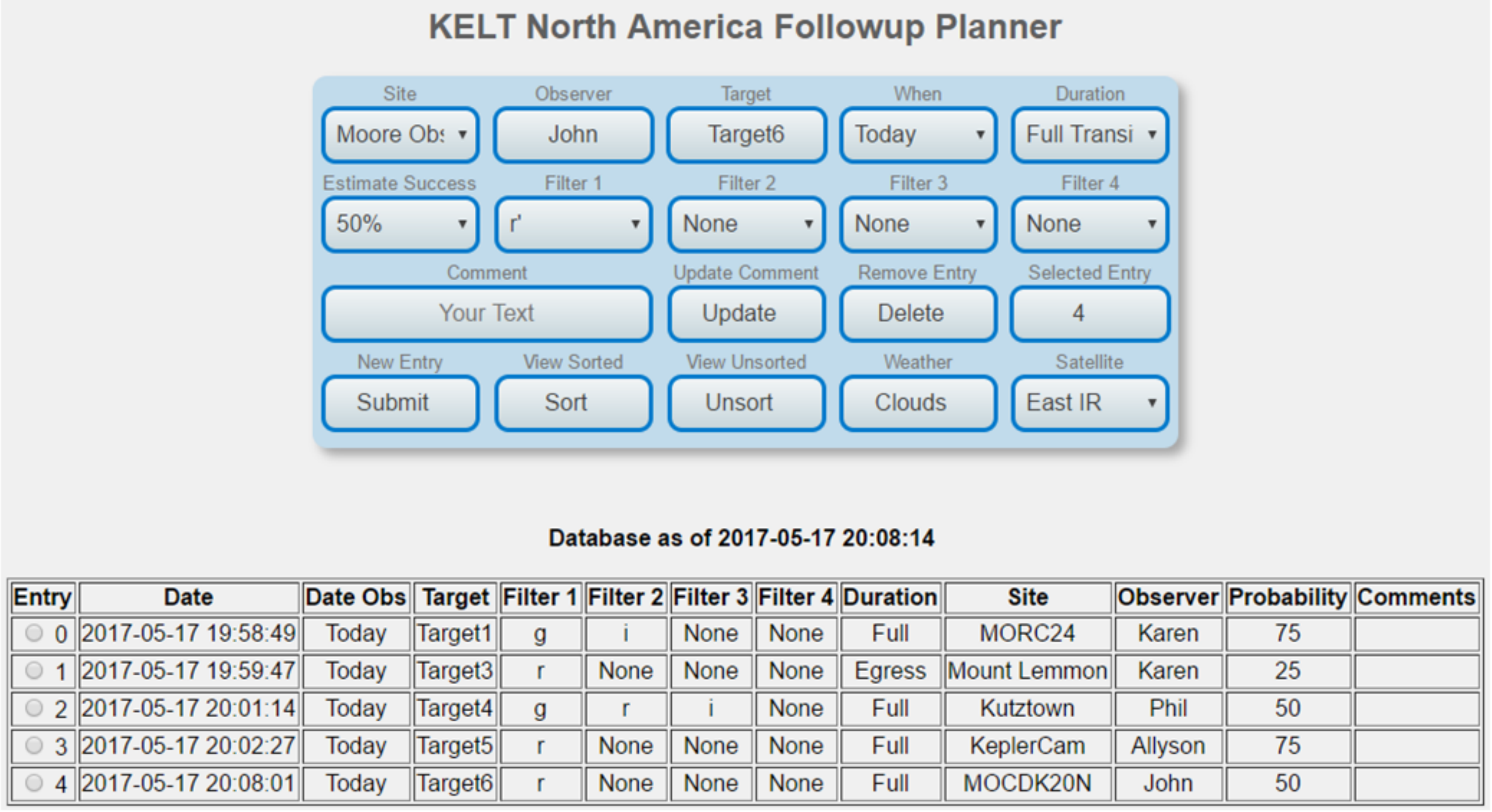}
\caption{The KELT-FUN Observations Coordinator for North America. Observers enter information about their own observing plans for specific targets on a given night to allow multiple observers to efficiently coordinate their observations and avoid unnecessary duplication of effort.
\label{fig:obscoordinate}}
\end{center}
\end{figure*}

A KELT Follow-up Observations Coordinator (KFOC) web tool is provided that allows the KELT-FUN team to optionally coordinate observations within a specific night to help avoid duplication of observations of the same object in the same filter, especially in cases where there are multiple simultaneous target events available for observation. KFOC is currently used by KELT-FUN members in the United States, where the density of observers is high (see Figure \ref{fig:kelt-fun_map}). As KELT-FUN expands, other regions will also be encouraged to use the site. The online interface of the tool is shown in Figure \ref{fig:obscoordinate}. Observers enter the target they plan to observe, the planned filter(s), the planned observational coverage of the event (full, ingress, egress, etc.), their site, name, and the probability of successful observations. Newly entered observations are then automatically propagated to all other users monitoring the website. 

\subsubsection{Reduction and Submission of Follow-up Photometry}

KELT-FUN team members calibrate their own images and extract differential photometry in preparation to submit results to the KELT Science Team. There is no requirement to use a specific software package to reduce data, but many KELT-FUN members use AstroImageJ (AIJ; \citealt{Collins:2017}) since it was developed out of the KELT-FUN effort, and support and training is readily available. 

Observers are asked to submit a short summary of their observations and a finder field indicating the target star, comparison stars, nearby stars searched for deep events, and any NEBs identified (if applicable). Observers also submit a light curve plot showing the target star light curve, the NEB light curve (if applicable), a sample of comparison star light curves, and a data table containing time of mid-exposure, differential photometry, photometric uncertainty, and any important detrending parameters. AIJ produces all of these data products in a format that is ready to submit with a single save operation. 

Observers are encouraged to submit results within 1-2 days of the observations to minimize the chance that another observer will duplicate observations of the same candidate, which may be a false positive. The submission data products are attached to an email and sent to a group email list that distributes the results to the KELT Science Team for analysis and to all other KELT-FUN members. We find that distributing results to all KELT-FUN members builds a sense of teamwork and camaraderie, and allows team members to learn observing strategies and data reduction techniques from each other.

\subsubsection{Use and Analysis of Follow-up Photometry}

When a KELT-FUN team member observes a candidate with photometry, it is their responsibility to extract a light curve for the target star, check nearby sources for variability, and contribute a brief report of the observation. A member of the KELT Science Team is then responsible for analyzing the newly-submitted observation, with additional context given by the KELT survey data and any past follow-up observations. At this stage, the KELT Science Team member will either expire the candidate or decide on an updated observing strategy. This decision and the reasoning behind it is then communicated to the full KELT-FUN and KELT Science teams.

For many reasons, the exact procedure for following up candidates differs from target to target. The utility of different types of observations and the order in which they are acquired depend on the target's observability, telescope resources, and the various scenarios that can most easily be revealed by a given type of observations. Here we describe a typical sequence, and explain how FPs are identified at various steps along the process. 

After identifying a transit candidate in the KELT survey data, a time-series photometric follow-up observation of the transit is requested. The most useful photometric observations will cover an ingress and/or egress and more than 50\% of the predicted duration, plus 30 minutes or more of pre-ingress and/or post-egress out of transit baseline. Once this observation is complete, the observer will reduce the data, extract a light curve for the target star, and also check the light curves of neighboring stars. Observers are expected to check all detected stars within a $3\arcmin$ radius, and to check nearby stars having a brightness comparable to or greater than the target star out to $6\arcmin$, to account for the possibility of the blending of extended PSF wings into the target star's aperture in the survey data. In practice, most NEBs are found within $1\arcmin - 2\arcmin~(\sim3-5$ KELT pixels) of the target star. After this data reduction and initial analysis, the observer will send their results to the full KELT-FUN and KELT Science teams, whereupon a member of the KELT Science Team will analyze the data further, and ultimately decide how to proceed.

If the observation reveals no variability on target, but a deep event in a nearby star, the candidate is expired as an NEB, so long as the neighbor is close enough to the target star on the sky to have caused the event that was identified in the survey data, the eclipse timing is consistent with the predicted ephemeris, and the event depth in the neighboring star is deep enough to exclude a transiting planet as the source of the signal. In some cases, a transit signal is detected in a neighboring star, but with a shallow depth consistent with itself being planetary in origin (\textit{e.g.} KELT-16b; \citealt{Oberst:2017}). Situations like this underscore the need for caution when analyzing follow-up observations, as to not hastily reject viable candidates. 

If the observation does show an event on the target star, the Science Team member will analyze the follow-up light curve and the KELT survey data together. This may lead to an improved ephemeris, and/or better knowledge of the depth of the event. With a more accurate depth measured from a follow-up light curve, the KELT Science Team member will estimate the size of the transiting body using a simple model. This model takes the transit depth and the stellar temperature as input (the effective temperature either estimated from archival broadband photometry, or from a previously acquired spectrum, when available), and calculates the radius of the transiting body. The model assumes the host star is at the zero-age main sequence (where there is a one-to-one correspondence between temperature and radius). If the transiting body is estimated to be larger than 2.5 Jupiter radii, the candidate is expired as an EB. Otherwise, the estimated companion size is consistent with planetary, and further follow-up observations are requested. 

Typically, the next step is to capture the transit in a different photometric filter, with the goal of measuring a full event in both a blue and a red passband that are reasonably well-separated in wavelength (\textit{e.g.} B and I). This allows us to test if the transit depth is chromatic, which is indicative of an EB or a BEB. We do not adopt a strict metric when deciding to expire a candidate based on filter-dependent transit depths, because light curve quality (noise, systematic trends, sky conditions, airmass, and telescope capabilities) can vary greatly between different observations. In general, differences in transit depths due to limb-darkening variations across different filters are not large enough to be detectable with the facilities involved in KELT-FUN, and so should not be the cause of detectable depth chromaticity. In practice, a difference of ~$\gtrsim$~5 mmag between the transit depth measured in different filters gives sufficient confidence to expire the candidate (so long as both light curves are of high quality). Whenever there is doubt, more data are requested.

Another photometric test for an EB scenario is to check for a possible odd/even transit/eclipse depth difference at twice the BLS-recovered period. For an EB, the BLS algorithm will often select half the orbital period. If doubling the BLS period results in a light curve with a primary and secondary eclipse of different depths, then the candidate is expired as an EB.

\subsection{Reconnaissance Spectroscopic Vetting}

The photometric follow-up observations are generally the first round of observations once candidates have been identified by the KELT candidate selection process, and for candidates that make it through the photometric observations, we then obtain spectroscopic observations. On occasion, a target that emerges from the vetting process described in \S \ref{sec:hum_vet} will be sent directly for spectroscopic observations. That happens when the KELT survey light curve provides an unambiguous sign of a transit-like feature with a reliable period, and there are no indications in images from the Digitized Sky Survey\footnote{https://archive.stsci.edu/cgi-bin/dss\_form} of any stars close enough to the target to have caused the detected transit signal after blending with the target. Spectroscopy may also be requested simultaneous with photometry due to scheduling reasons and for long period transits.

Spectroscopic observations are not organized in the same way as photometric observations. There are only a few members of KELT-FUN with spectroscopic capabilities with the spectroscopic resolution and signal-to-noise ratio (SNR) requirements needed to rule out FP spectroscopically. Lists of candidates slated for spectroscopic observations are compiled by members of the KELT Science Team and sent to individual observers with the available resources. Reconnaissance spectroscopic observations aimed at efficiently identifying astrophysical FPs are performed with the Tillinghast Reflector Echelle Spectrograph (TRES; \citealt{Szentgyorgyi:2007, Furesz:2008}), on the 1.5 m telescope at the Fred Lawrence Whipple Observatory (FLWO) on Mt. Hopkins, Arizona\footnote{http://www.sao.arizona.edu/FLWO/whipple.html}, and with the Wide Field Spectrograph (WiFeS;  \citealt{Dopita:2007}) on the Australian National University (ANU) 2.3~m telescope at Siding Spring Observatory in Australia\footnote{
http://rsaa.anu.edu.au/observatories/telescopes/anu-23m-telescope}. These observations usually follow the procedure laid out in \citet{Latham:2009}. An initial observation taken at quadrature is obtained and cross correlated against a library of synthetic spectral templates to estimate the stellar atmospheric properties, including \teff, \vsini, \loggstar; if the host star is deemed to be evolved, the candidate is usually rejected as the transiting companion would not be of planetary radius. The cross correlation function is examined for signatures of contamination due to stellar binary or blended companions. Candidates that pass the initial inspection then receive additional observations timed at the opposite quadrature to check for large RV variations ($\gtrsim 1 \mathrm{km\,s}^{-1}$) that may be induced by stellar-mass companions. The presence of multiple lines or large velocity variations alone is not necessarily a reason to expire a planet candidate (e.g. KELT-1~b; \citealt{Siverd:2012}). If the line movements or RV variations are not in phase with the photometric period, the candidate could still be a valid planetary system (e.g. KELT-19~Ab; \citealt{Siverd:2018}). For the Southern candidates, we also have the unique capability of examining the candidates and surrounding stars with the WiFeS \citep{Bayliss:2013} integral field spectrograph. With WiFeS, we are able to simultaneously obtain spectra for stars that are nearby (and blended in KELT photometry) with the KELT target star. We can therefore search for stellar eclipsing binaries that may be blended in the KELT photometry. Typically, this only requires a few spectroscopic measurements taken at quadrature as predicted by the KELT transit ephemeris.

\subsection{Confirmation and Final Vetting of Candidates}

At this point, if a candidate cannot be ruled out as an FP by any of the aforementioned tests, it is potentially a genuine transiting exoplanet. Typically, more photometric observations are requested to refine the ephemeris, and transit depth and shape, while more high-precision spectroscopic follow-up observations are requested to measure the RV orbit of the system (constraining the planetary mass and orbital eccentricity), and to improve our knowledge of the stellar parameters. In the cases of slowly rotating host stars, we often can obtain precise masses of the planetary companions with high resolution, high signal-to-noise RV measurements, at the 3-30\,$\mathrm{m\,s}^{-1}$ level, across the orbital phase of the system. In addition, we also examine the line bisector spans to confirm the detected RV orbits are not induced by blended background binaries \citep{Mandushev:2005}.

If the host star is found to be a rapid rotator with \vsini\ greater than about 40 km s$^{-1}$, then it is often not possible to precisely measure the RV orbit due to the rotational broadening of the spectral features (and, since rapidly rotating stars tend to be hotter, a lack of sufficient spectral features by which RV measurements are made). In these cases, an upper limit can sometimes be placed on the mass of the transiting body, perhaps excluding an EB scenario consisting of a relatively massive main sequence primary and a low mass (but still stellar) secondary.  If the transit depth indicates a planetary sized companion, such a candidate can be confirmed through a Doppler tomographic \citep[DT, see][]{Collier:2010b} analysis (up to rotational speeds of 
\vsini$\approx$ 200 km s$^{-1}$). DT analysis can confirm that the cause of the photometric signal is indeed a planet-sized body transiting the rapidly-rotating target star, ruling out a BEB scenario. This is an integral part of the KELT discovery strategy, since a significant fraction of our candidates (and also a large fraction of our confirmed planet discoveries) have hot, rapidly rotating host stars \citep{Bieryla:2015}. 

If a planet candidate has passed all of the aforementioned cuts, adaptive optics (AO) observations are requested. AO observations can reveal the existence and flux of projected nearby stars. Accounting for the contaminating flux from these nearby stars (if they exist) results in improved parameters of the planetary system. If a projected nearby companion is not detected in the AO data, useful limits on the existence of potential companions can be placed as a function of magnitude difference and projected separation from the target star. 

Finally, if a planetary mass companion is confirmed from photometric time-series imaging, AO imaging, and RV and/or DT analysis, the candidate is promoted to a confirmed exoplanet and the discovery publication process begins. 

The various FP scenarios are usually confidently classified as such. However, determining transit candidate FAs can be more difficult. If a follow-up light curve does not show a transit-like event at the predicted time, the Science Team member must consider the quality of both the follow-up light curve and the KELT survey data. If there truly is an event on target at the predicted time, but is of a shallow depth, then the event may evade detection if the scatter in the follow-up light curve is similar to or greater than the transit depth, or if systematic effects (\textit{e.g.} a trend with airmass, or deteriorating sky conditions) dominate. In this situation, additional observations are scheduled, with a request that the target only be observed if a high SNR is achievable, as observations with high scatter can neither rule out nor confirm the presence of a shallow transit. These situations must be dealt with carefully, as to not expire viable candidates with events that are difficult, but not impossible, to detect (\textit{e.g.} KELT-11b, with a 0.25\% transit depth; \citealt{Pepper:2017}).  
Sometimes, the particular configuration of a candidate system can lead to a planetary confirmation even with ambiguity regarding the host star (\textit{e.g.,} NGTS-3Ab, a binary system with a transiting planet; \citealt{Gunther:2018}).  If a follow-up light curve can confidently rule out the existence of a transit at the predicted time, then an alternative ephemeris will be explored (if any viable alternatives exist). This may correspond to another strong peak found by the BLS algorithm, or an ephemeris with twice the original BLS-determined period, but half a phase away from what has already been covered by a follow-up observation. Another photometric observation at the new ephemeris is then acquired. If, after two (or more, if needed) observations, there is no evidence for a transit in the follow-up data, the candidate will typically be expired as an FA. This process is somewhat subjective, so an FA classification could still be incorrect if the ephemerides derived from the KELT data lack the precision needed to predict the transit center time within approximately the duration of the transit event at the epoch of the follow-up observations. It is at the discretion of the Science Team member to decide if a suspected FA candidate is worth any additional follow-up resources, or if those resources are better spent on candidates that are more likely to yield results.  We therefore stress that, for this reason, and many others, we do not claim that our FP catalog is complete in any sense.  

\section{The KELT False Positive Catalog}

We present the results of 1,128 KELT-FUN FP detections in machine readable catalog format to help minimize duplicate follow-up observation efforts by current and future transiting planet wide-field surveys such as TESS. 

\subsection{False Positive Categories}

KELT FPs are classified into nine types that are organized into two broad categories -- Spectroscopic FPs and Photometric FPs -- as shown in Table \ref{tab:catalogsummary}. If only spectroscopic or photometric follow-up observations were obtained before confirming a candidate as an FP, one of the corresponding categories was assigned. If the FP was detected in both spectroscopic and photometric follow-up observations, the observation that provided the highest confidence in an FP categorization was used to assign an FP category. When both photometric and spectroscopic observations support the same underlying astrophysical configuration with similar confidence, we generally arbitrarily assigned a spectroscopic category. In addition to the confirmed FPs reported here, we classified about 450 KELT candidates as FAs. Most or all of the FAs are believed to have been caused by spurious signals in the KELT data, and thus are not included in the KELT FP catalog.  Nevertheless, as stressed above, we cannot be completely confident that some of the candidates that we designated as FAs (and thus are not included in our catalog) are, in fact, astrophysical FPs.  Again, we make no claims as to the completeness of our catalog over any region of parameter space.

Spectroscopically detected FPs are separated into four categories labeled RV0, Giant, SB1, and SB2 in the catalog. 
The RV0 category is assigned if a photometric event has been confirmed by KELT-FUN, but no significant RV variation is detected in spectroscopic follow-up, at a level that rules out the presence of a giant planet at the nominally detected period. The Giant category is assigned if the target star is spectroscopically identified as a giant star which was not identified and removed by the reduced proper motion cut and there is no detected velocity variation. Eclipses of giant stars detected by KELT are likely caused by stellar companions since a planetary transit of a giant star would, in general, be too shallow to be detectable by KELT.
The SB1 category is assigned if two or more spectra show a single-lined stellar spectrum with an RV semi-amplitude that is too large to be consistent with a planetary companion (${\rm K}\gtrsim 1~ \mathrm{km\,s}^{-1}$) and the velocities are not inconsistent with the photometric ephemeris. 
Finally, the SB2 category is assigned if one or more spectra show a multi-lined composite spectrum that is consistent with multiple blended stars and an RV variation that is consistent with the photometric ephemeris or is too large to be consistent with a planetary companion. 

Regarding the SB2 category, when we detect a composite spectrum we set it aside and do not invest additional telescope time to determine an orbit.  Getting agreement with both the period and epoch between the nominal photometric ephemeris and an orbital solution would be the only way to prove that eclipses of two stars are the source of the shallow transit-like dips.  Early in the history of the project, when there were fewer candidates, we did follow up many eclipsing binaries to show that stellar eclipses explained the light curves \citep[see][for details]{Latham:2009}. These early observations showed that quite often the photometric ephemeris had the period wrong by a factor of 2, and occasionally by more exotic factors. In principle there could still be a planet around one of the stars in a system with composite spectra, but it will be almost impossible to say anything reliable about the mass and radius without an inordinate amount of additional observations and effort.  Thus, while technically these cases are not necessarily false positives, we regard them as FPs for all intents and purposes of the KELT survey.

Photometric false positives are separated into five categories labeled EB1, EB2, BEB, Variable, and NEB in the catalog. The EB1 category is assigned if the deblended transit depth in the follow-up photometry is too deep, relative to the host star's estimated radius, to be consistent with a transiting planet companion. The category EB2 is assigned if even numbered orbits have a depth different from the odd numbered orbits, indicating primary and secondary eclipses of an EB or blended EB system. If significantly different transit depths are measured in blue and red filters, the BEB category is assigned. Eclipses showing depth chromaticity can be caused by EBs blended with the target star or a hierarchical stellar system in the follow-up observations photometric aperture. This can also be caused by an unblended eclipsing binary system consisting of two stars with different surface temperatures, and where the light from the secondary is not negligible compared to the primary. In some cases variable stars that are not in eclipsing systems cause a KELT detection. We categorize those FPs as simply Variable. The most common photometric FPs result from transiting candidate host stars with nearby eclipsing binary systems, or NEBs, that are blended with the target star in the KELT aperture, but are not blended in the follow-up aperture.

\begin{table}
\centering
\footnotesize
\caption{KELT False Positives by Category\textbackslash Type}
\label{tab:catalogsummary}
\setlength\tabcolsep{3.0pt}
\begin{tabular}{lllr} \hline \hline
\multicolumn{2}{l}{Category~~~~~~~~~~~~~~Type} & Description & Total \\ \hline
Spectroscopic FPs & & &   \\
~~~SB1   \dotfill & 1    & Single-lined binary (RV $\gtrsim 1~ \mathrm{km\,s}^{-1}$)           & 307   \\
~~~SB2   \dotfill & 2    & Multi-lined binary                       & 140   \\
~~~RV0   \dotfill & 3    & No significant RV detected               & 13    \\
~~~Giant \dotfill & 4    & Spectroscopic Giant                      & 29    \\
Photometric FPs   &  &   &   \\
~~~EB1  \dotfill & 5     & Too deep in follow-up                    & 130   \\
~~~EB2  \dotfill & 6     & Different primary \& secondary depths    & 25    \\
~~~BEB  \dotfill & 7     & Blend in follow-up aper. (chromaticity)  & 90    \\
~~~Variable \dotfill & 8 & Variable star caused KELT detection      & 16    \\
~~~NEB  \dotfill & 9     & Nearby EB (Blend in KELT aperture)       & 378   \\\cline{4-4} 
\rule{0pt}{2.5ex}  Total             &  &   & 1128   \\ \hline    
\end{tabular}
\end{table}

\subsection{Information in the Catalog}\label{sec:catfields}

The data fields provided in the catalog are described in Table \ref{tbl:catdatanames}. Certain fields may be empty due to unavailable data or non-applicable fields for certain FP types. In addition to the FP classification, candidate host star IDs, including the KELT, 2MASS, and TESS Input Catalog (TIC) IDs, equatorial, Galactic, and ecliptic coordinates, and V magnitude are included. 

For the associated candidate transit event measured from the KELT light curves, the transit center time and uncertainty, orbital period and uncertainty, transit duration, and transit depth as measured in the KELT aperture are provided. For SB1 type FPs, the RV semi-amplitude is provided, if available, and for EB1 type FPs, the EB depth is included, if available.

For NEB type FPs, we provide information related to the nearby eclipsing system, if available. In some cases, the right ascension and declination (J2000) of the target is provided, but for most NEBs, the approximate distance and direction from the candidate host star to the NEB star is provided. A flag indicating that the distance is approximate is set if the precise measured distance was not readily available. We also provide the depth of the NEB, as measured from the follow-up photometry, if available. Two flags are associated with NEB eclipse depths. The first flag indicates that the depth was estimated by eye from a plot, and the second flag indicates that the depth was a lower limit due to the follow-up photometry not including both pre- or post-eclipse and mid-eclipse coverage. Finally, the date and filter band of the NEB follow-up observations is provided, if available.

\begin{table}
\footnotesize
\caption{Description of False Positive Catalog Data Columns}\label{tbl:catdatanames}
\label{columns}
\setlength\tabcolsep{3.0pt}
\begin{tabular}{ll}
\hline
\hline
Column Name & Description \\ \hline
KELT\_ID & KELT Survey candidate ID \\
2MASS\_ID & Two Micron All-Sky Survey ID \\
TIC\_ID & TESS Input Catalog ID \\
In\_CTL & Flag: star is in TESS Candidate Target List \\
TESS\_priority & Priority from TESS Input Catalog \\
FP\_type\_name & False positive type name \\
FP\_type & False positive type number \\
RA\_hours & Right ascension in hours (J2000)\\
RA\_deg & Right ascension in degrees (J2000)\\
RA\_hms & Right ascension  in sexagesimal (J2000)\\
Dec\_dms & Declination in sexagesimal (J2000)\\
Dec\_deg & Declination in degrees (J2000)\\
Galactic\_long & Galactic longitude in degrees \\
Galactic\_lat & Galactic latitude in degrees \\
Ecliptic\_long & Ecliptic longitude in degrees \\
Ecliptic\_lat & Ecliptic latitude in degrees \\
Vmag & V magnitude of star \\
Tc & Transit center time in \bjdtdb \\
Tc\_err & Uncertainty in Tc \\
Period\_days & Period of transit in days \\
Period\_err & Uncertainty in period \\
Duration\_hrs & Duration of transit in hours \\
KELT\_depth\_mmag & Depth of transit in KELT aperture in mmag \\
EB\_K\_km/s & RV semi-amplitude of EB in km/s \\
EB\_depth\_mmag & Depth of EB transit in mmag \\
NEB\_RA & Right ascension of nearby eclipsing binary \\
NEB\_Dec & Declination of nearby eclipsing binary \\
NEB\_dist\_text & Distance from target star to NEB \\
NEB\_dist\_arcsec & Numeric distance from star to NEB in arcs \\
NEB\_dist\_is\_approx\_flag & Flag: distance to NEB is approximate \\
NEB\_direction & Direction from star to NEB \\
NEB\_depth\_text & Depth of NEB transit \\
NEB\_depth\_percent & Numeric depth of NEB transit in mmag \\
NEB\_depth\_is\_approx\_flag & Flag: NEB depth is approximate \\
NEB\_depth\_is\_lower\_limit & Flag: NEB depth is a lower limit \\
NEB\_obs\_epoch\tablenotemark{a} & Date NEB was observed \\
NEB\_obs\_filter & Filter used to observe NEB \\ \hline
\end{tabular}
\tablenotetext{a}{Note that on some occasions an NEB was observed on multiple nights. The main reason is that additional observations took place prior to the submission of the results of an earlier observation.}
\end{table}

\subsection{The Catalog Data}

The catalog data are provided in machine readable tabular format in the online version of this article. The data are organized as specified in \S \ref{sec:catfields}. An example of the data provided for all FP types is shown in Table \ref{tbl:catmain}. An example of each of the nine FP types is provided. Table \ref{tbl:catebvalues} shows the two data columns that include RV semi-amplitude for some SB1 systems and eclipse depths for some EB1 systems. Finally, Table \ref{tbl:nebcatvalues} shows four examples of additional data that are included for NEB type FPs, if available. The values in all three tables are included in a single line of the catalog. The KELT ID column is repeated in the table panels for clarity and is not repeated in the catalog. 

\begin{figure}
\begin{center}
\includegraphics[width=1.0\linewidth,trim=2.0mm 0.0mm 0mm 0mm,clip]{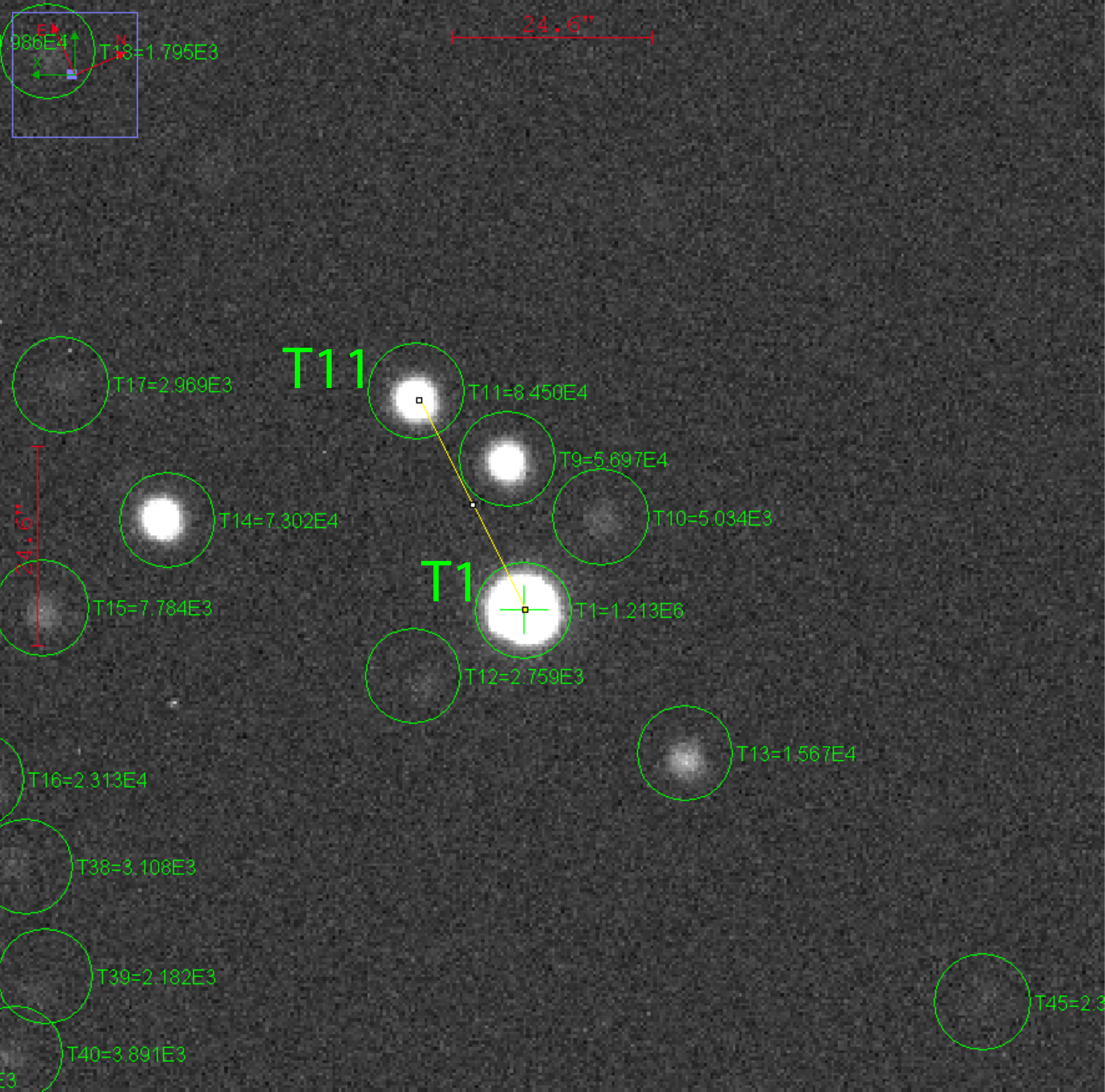}
\caption{Example of a sky image of a target field with an NEB, produced by AIJ. The target star aperture is marked T1. The additional sources encircled with a green aperture were checked for NEBs. In this case, the star in the aperture marked T11 was determined to be an NEB. The yellow line shows the direction from the target star to the NEB. The red text and bars show the horizontal and vertical scales of the image. Similar figures are provided for most NEB type FPs at the KELT False Positive FilterGraph Portal.
\label{fig:NEBfield}}
\end{center}
\end{figure}

\begin{figure}
\begin{center}
\includegraphics[width=1.0\linewidth,trim=0mm 0.0mm 0mm 0mm,clip]{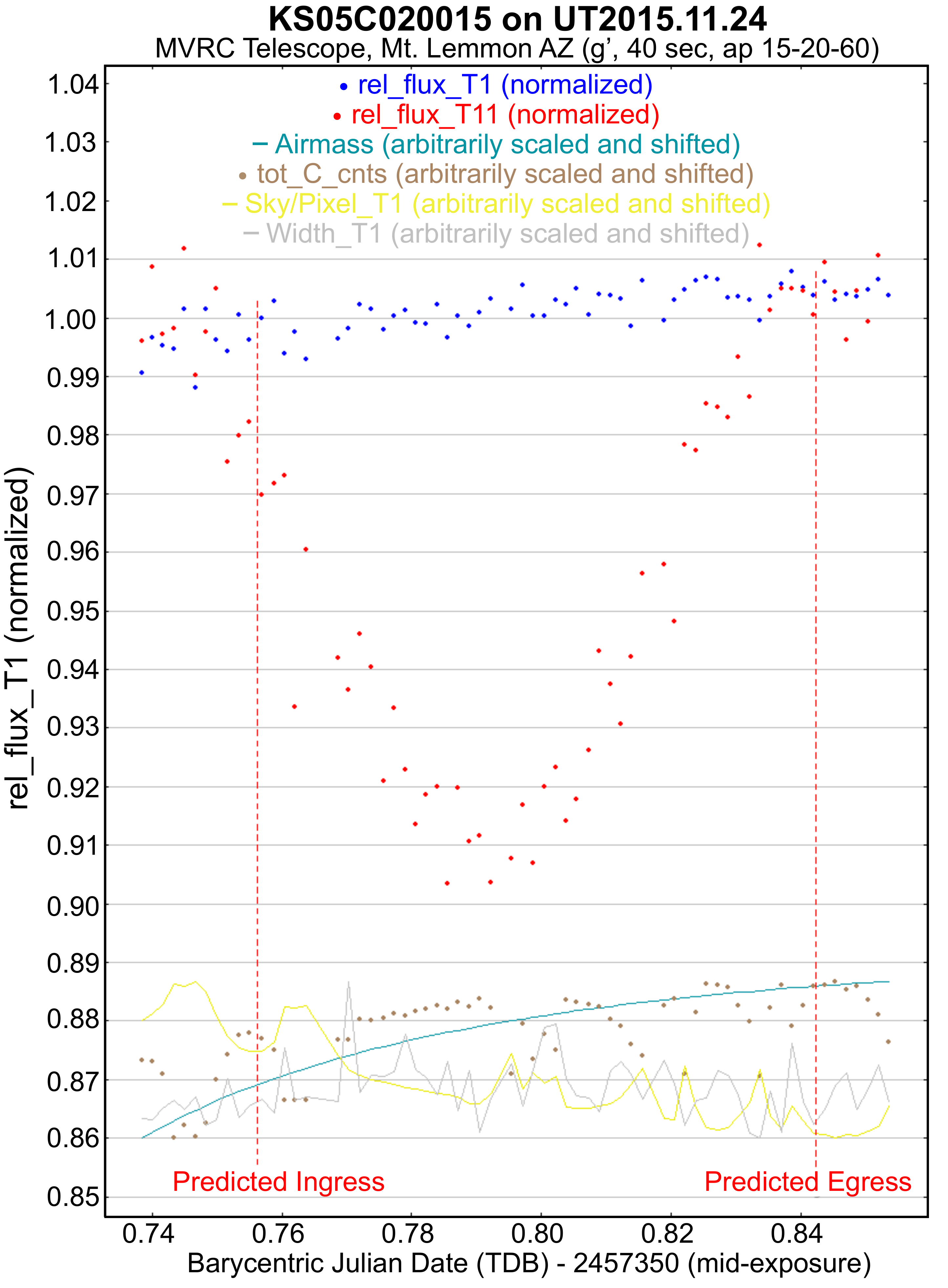}
\caption{Example of a light curve plot of an NEB, produced by AIJ. The normalized flux of the candidate target star is shown as blue dots. The NEB light curve is shown as red dots. The KELT-predicted ingress and egress times are shown as red vertical dotted lines. Arbitrarily scaled and shifted airmass (inverted), total comparison star counts, sky-background, and average FWHM are shown as teal lines, brown dots, yellow lines, and grey lines, respectively. Similar figures are provided for most photometric FPs at the KELT False Positive FilterGraph Portal.
\label{fig:NEBlightcurve}}
\end{center}
\end{figure}

The catalog is also available through a FilterGraph \citep{Burger:2013} portal\footnote{https://filtergraph.com/kelt\_false\_positive\_catalog} for ready access to catalog data and plotting. The FilterGraph portal also includes links to plots of field images which may show the locations of comparison stars used for the differential photometry and the position of an NEB relative to the candidate host star, if applicable. Figure \ref{fig:NEBfield} shows an example NEB field image. Also included in the FilterGraph portal are links to light curve plots showing the target star light curve, and if applicable, the NEB light curve. Figure \ref{fig:NEBlightcurve} shows an example NEB light curve plot. For EB1 type FPs, links are provided to plots of the phased RV data and best fit orbital model, when available.

\begin{table*}
\footnotesize
\caption{KELT False Positive Catalog (Data Common to All False Positives)}
\label{tbl:catmain}
\begin{tabular}{ccccccccc}
\hline
\hline
KELT ID & 2MASS ID & TIC ID & In CTL & TESS Priority & RA (J2000)& RA (J2000) & RA (J2000) & Dec (J2000)  \\
 &  &  &  &  & (h:m:s) & (hours) & (degrees) & (d:m:s)    \\ \hline
KJ06C001078 & J06593615+0104056 & 237853540 & 1 & 0.000192674512176
 & 06:59:36.10 & 6.9933611 & 104.9004167 & 01:04:05.60    \\
KJ06C059566 & J07022053+0420573 & 291308749 & 1 & 0.000921388339681 & 07:02:20.50 & 7.0390278 & 105.5854167 & 04:20:57.30   \\
KS14C001431 & J19215239+0706085 & 132022468 & 1 & 0.000557041092605 & 19:21:52.41 & 19.3645579 & 290.4683687 & 07:06:08.34    \\
KS13C017379 & J18263815-0838339 & 385835154 & 1 & 0.000479365808543 & 18:26:38.16 & 18.4439339 & 276.6590090 & -08:38:33.91    \\
KS13C018108 & J17362073+0955340 & 277626665 & 1 & 0.00078259577339 & 17:36:20.74 & 17.6057597 & 264.0863961 & 09:55:33.90   \\
KS05C044312 & J06160057+0619299 & 274235078 & 1 & 0.000835138112045 & 06:16:00.58 & 6.2668265 & 94.0023979 & 06:19:29.96  \\
KJ06C000533 & J06544012+0643268 & 235380067 & 1 & 0.000781927886478 & 06:54:40.10 & 6.9111389 & 103.6670833 & 06:43:26.70   \\
KS19C02564 & J02134607-4146319 & 138735221 & 1 & 0.000825362797298 & 02:13:46.10 & 2.2294722 & 33.4420833 & -41:46:31.80   \\
KJ06C019953 & J08001402+0706385 & 320538316 & 1 & 0.000892102663055 & 08:00:14.00 & 8.0038889 & 120.0583333 & 07:06:38.50  \\
\vdots & \vdots & \vdots & \vdots & \vdots & \vdots & \vdots & \vdots & \vdots \\ \hline
 \\
 \\

\hline
\hline
KELT ID & Dec (J2000) & Galactic & Galactic & Ecliptic & Ecliptic & Vmag & Tc & Tc Err  \\
 & (degrees) & Longitude & Latitude & Longitude & Latitude &  & (${\rm BJD_{TDB}}$) & (Days)  \\ \hline
KJ06C001078 & 1.0682222 & 212.912 & 2.3034 & 106.0459167 & -21.5448889 & 7.94 & 2457048.96 & 0.26  \\
KJ06C059566 & 4.3492500 & 210.295 & 4.40377 & 106.3809444 & -18.2081389 & 11.9 & 2457061.345 & 0.01   \\
KS14C001431 & 7.1023156 & 63.1132 & 7.28434 & 293.3507778 & 28.8974167 & 8.39 & 2457248.36 & 0.0041  \\
KS13C017379 & -8.6427529 & 26.0371 & 3.41557 & 276.8049722 & 14.6393889 & 11.24 & 2457197.443 & 0.013  \\
KS13C018108 & 9.9260830 & 31.2266 & 20.1281 & 263.0319722 & 33.2224444 & 12.3 & 2457197.691 & 0.0058  \\
KS05C044312 & 6.3249894 & 233.617 & -19.0231 & 94.1613333 & -17.0565000 & 11.56 & 2457344.24 & 0.0056   \\
KJ06C000533 & 6.7240833 & 273.735 & -23.7269 & 104.1327222 & -16.0471944 & 7.22 & 2457058.462 & 0.0034  \\
KS19C02564 & -41.775500 & 155.554 & -49.7563 & 10.2086944 & -50.7795833 & 9.59 & 2456601.696 & 0.016  \\
KJ06C019953 & 7.1106944 & 209.427 & 20.7299 & 120.6959444 & -13.1830278 & 11.28 & 2457058.008 & 0.0026  \\
\vdots & \vdots & \vdots & \vdots & \vdots & \vdots & \vdots & \vdots \\ \hline
 \\

\hline
\hline
KELT ID & Period & Period Err & Duration & KELT Depth & FP Type & Type \\
 & (days) & (days) & (hours) & (mmag) & Name &  \\ \hline
KJ06C001078 & 18.7956 & 0.0041 & 13.44 & 9.13 & SB1 & 1 \\
KJ06C059566 & 1.292033 & 0.000011 & 2.7168 & 14.7 & SB2 & 2 \\
KS14C001431 & 1.8590175 & 0.000007 & 2.347 & 6.91 & RV0 & 3 \\
KS13C017379 & 12.04464 & 0.00017 & 5.352 & 21.1 & Giant & 4 \\
KS13C018108 & 3.587427 & 0.00002 & 3.696 & 18 & EB1 & 5 \\
KS05C044312 & 1.2119727 & 0.000006 & 2.062 & 16.1 & EB2 & 6 \\
KJ06C000533 & 0.32028678 & 0.000001 & 2.22 & 11.68 & BEB & 7 \\
KS19C02564 & 1.019386 & 0.000018 & 1.75 & 5.2 & Variable & 8 \\
KJ06C019953 & 0.6348532 & 0.0000016 & 1.5624 & 16.36 & NEB & 9 \\
\vdots & \vdots & \vdots & \vdots & \vdots & \vdots & \vdots \\
\hline
\end{tabular}
\end{table*}

\begin{table}
\centering
\footnotesize
\caption{KELT False Positive Catalog (Additional EB Data)}
\label{tbl:catebvalues}
\begin{tabular}{ccc}
\hline
\hline
KELT ID & EB K & EB Depth \\
 & km/s & (mmag) \\ \hline
KJ06C001078 & 28 &  \\
KJ06C001172 & 15 &  \\
KS13C018108 &  & 180 \\
KS14C005429 &  & 70 \\ \hline
\end{tabular}
\end{table}

\begin{table*}
\centering
\footnotesize
\caption{KELT False Positive Catalog (Additional NEB Data)}
\label{tbl:nebcatvalues}
\begin{tabular}{cccccccccccc} 
\hline
\hline
KELT ID & NEB RA & NEB Dec & NEB Dist & NEB Dist & NEB Dist is & NEB Dir \\
 & (J2000) & (J2000) & (text) & (arcsec) & Approx. Flag &  \\ \hline
KS36C077636 & 17:15:45 & -65:14:01 & 68" & 68 & 0 & SSW \\
KS27C034425 & 21:19:04 & -63:52:09 & 64" & 64 & 0 & SE \\
KS36C007691 & 17:32:35 & -41:27:10 & 7" & 7 & 0 & NNE \\
KS34C011419 & 9:10:54 & -53:55:53 & 21" & 21 & 0 & S \\
\vdots & \vdots & \vdots & \vdots & \vdots & \vdots & \vdots \\ \hline
 \\

\hline
\hline
KELT ID & NEB Depth & NEB Depth & NEB Depth is & NEB Depth is & NEB Obs & NEB Obs \\
 & (text) & (\%) & Approx. Flag & Lower Limit Flag & Date & Filter \\ \hline
KS36C077636 & $\sim$37\% & 37 & 1 & 0 & 20160825 & GG \\
KS27C034425 & 28\% & 28 & 0 & 0 & 20160912 & GG \\
KS36C007691 & 14\% & 14 & 0 & 0 & 20160917 & GG \\
KS34C011419 & $\sim$40\% & 40 & 1 & 0 & 20161220 & GG \\
\vdots & \vdots & \vdots & \vdots & \vdots & \vdots & \vdots \\ \hline
\\
\end{tabular}
\end{table*}

\section{Discussion}

The particular distribution of FP types for a given transit survey depends on many factors, including the survey design, the pixel scale, the photometric precision, pre-selection of target stars, and especially the sequence of follow-up observations. Consistent with other surveys, SB1s and NEBs are the dominant types of false positives for KELT.

Since KELT-FUN has significantly more photometric resources than spectroscopic resources, photometric follow-up is generally pursued first, except for long-period ($\mathrm{P}\gtrsim10$\,d), or other high value targets. Because of our photometry-first approach, there are more photometric FPs in the catalog than there would be if spectroscopy-first were employed. This demonstrates that for future wide-field transit surveys such as TESS, prioritizing relatively low-cost photometric observations, which can be conducted by facilities of all sizes, over more limited, and usually more expensive, spectroscopic observations can effectively reduce the workload on the more precious spectroscopic resources. This is especially the case when there is an extensive network of telescopes with apertures smaller than 1m, which will not be able to obtain precision RV, but which can reliably obtain sub-1\% photometry with seeing-limited angular resolution.

\begin{figure}
\begin{center}
\includegraphics[width=1.0\linewidth,trim=0.0mm 0.0mm 0.0mm 0.0mm,clip]{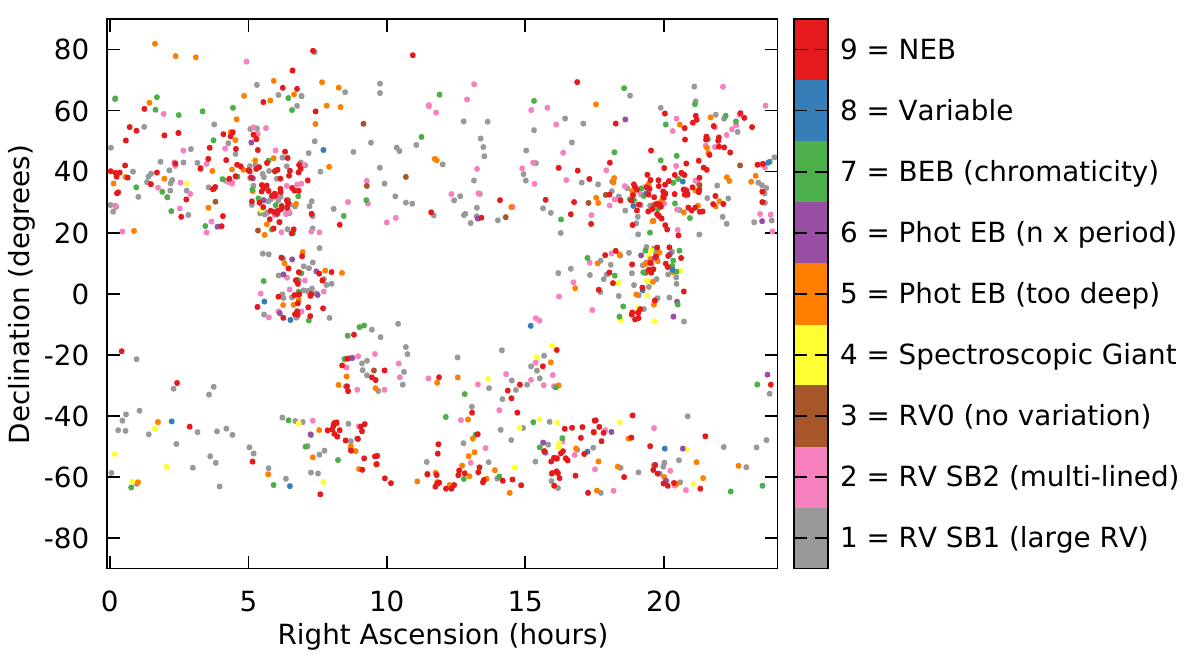}
\caption{Sky Location of all KELT False Positives. Symbol color represents FP type. The general regions of KELT sky coverage that have been followed-up are obvious and generally correspond to the KELT fields with the most data. Note that there are more FPs in the northern hemisphere than in the southern hemisphere because KELT-North has been running $\sim 5$ years longer than KELT-South. Also note the higher density of NEBs in the crowded galactic plane. The figure was created at the KELT False Positive FilterGraph portal.
\label{fig:FPs_location}}
\end{center}
\end{figure}

Figure \ref{fig:FPs_location} shows the sky location of all of the FPs included in the KELT FP Catalog. Symbol color represents FP type. The general regions of KELT sky coverage that have been followed-up are obvious and generally correspond to the KELT fields with the most data. Note that there are more FPs in the northern hemisphere than in the southern hemisphere because KELT-North has been running $\sim 5$ years longer than KELT-South. Also note the higher density of NEBs in the crowded galactic plane. The overall dominance of the NEB and SB1 types is easily visualized from the high density of red and grey symbols, respectively.

\begin{figure}
\begin{center}
\includegraphics[width=1.0\linewidth,trim=0.0mm 0.0mm 0.0mm 0.0mm,clip]{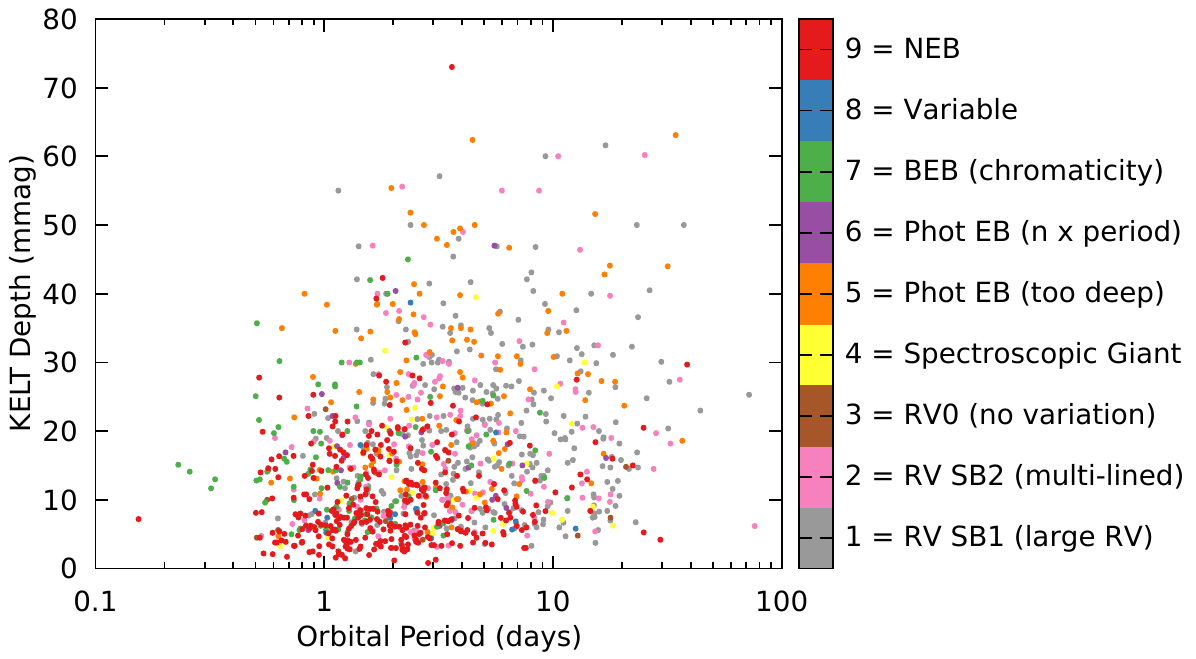}
\caption{KELT Depth vs. Period. Symbol color represents FP type. Photometric NEBs dominate at periods less than $\sim 10$ days since photometry-first is generally pursued for those candidates while SB1s dominate the longer period KELT detections. The figure was created at the KELT False Positive FilterGraph portal. 
\label{fig:FPs_period_depth}}
\end{center}
\end{figure}

\begin{figure}
\begin{center}
\includegraphics[width=1.0\linewidth,trim=0.0mm 0.0mm 0.0mm 0.0mm,clip]{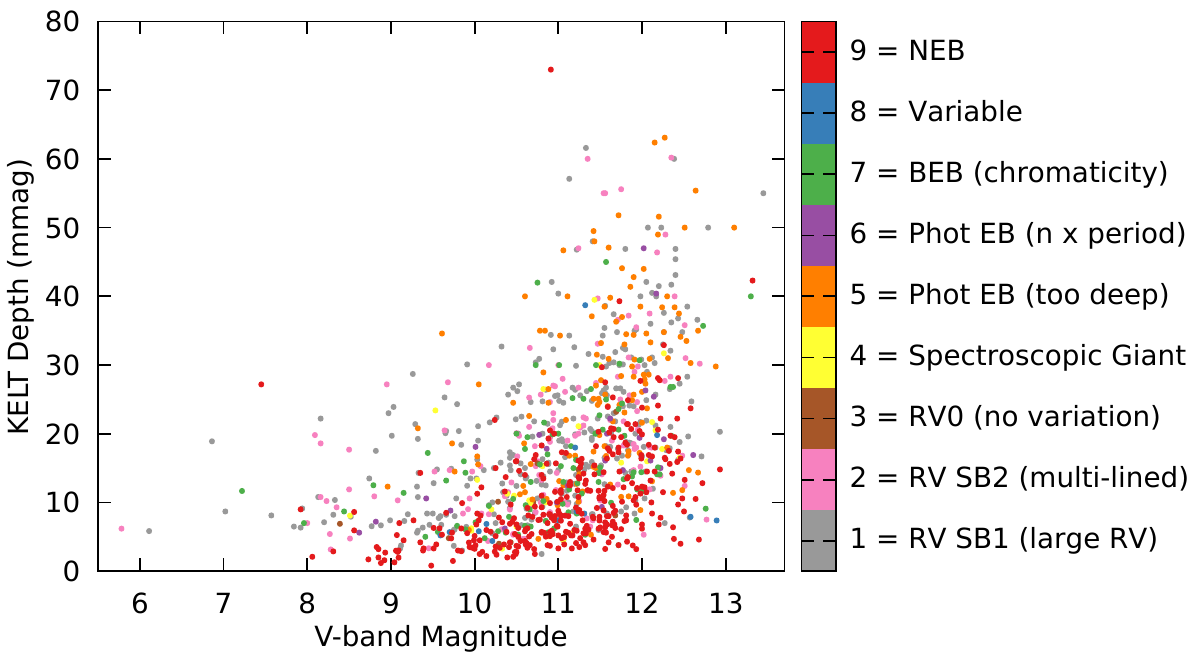}
\caption{KELT Depth vs. V-band Magnitude. Symbol color represents FP type. Photometric NEBs dominate at depths less than $\sim10$~mmag. The figure was created at the KELT False Positive FilterGraph portal. 
\label{fig:FPs_depth_vmag}}
\end{center}
\end{figure}

Figure \ref{fig:FPs_period_depth} displays all KELT FPs as KELT-detected depth vs. period. Symbol color represents FP type. Photometric NEBs dominate at periods less than $\sim 10$ days since photometry-first is generally pursued for those candidates while SB1s dominate the longer period KELT detections. Figure \ref{fig:FPs_depth_vmag} displays all KELT FPs as KELT-detected depth vs. V-band magnitude. Symbol color represents FP type. Photometric NEBs dominate at depths less than $\sim10$~mmag.

\begin{figure}[!hb]
\begin{center}
\includegraphics[width=1.0\linewidth,trim=0.0mm 0.0mm 0.0mm 0.0mm,clip]{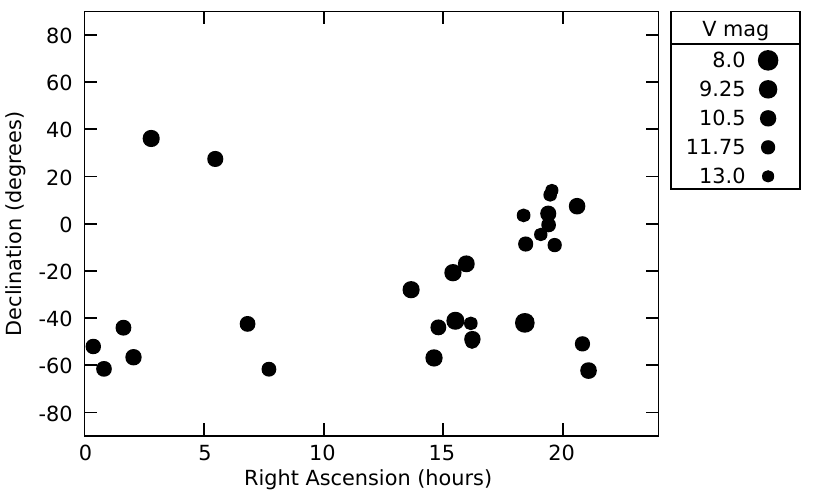}
\caption{Spectroscopic Giant False Positives. The symbol size represents V-band magnitude. Note that despite the larger number of FPs in the northern hemisphere, most of the spectroscopic giant FPs are in the southern hemisphere. The figure was created at the KELT False Positive FilterGraph portal.
\label{fig:FPs_giants}}
\end{center}
\end{figure}

Figure \ref{fig:FPs_giants} shows the sky location of all spectroscopic giant FPs. The symbol size represents V-band magnitude. These are giant stars that passed the reduced proper motion cut and made it into the KELT input catalog. Note that despite the larger number of FPs in the northern hemisphere, most of the spectroscopic giant FPs are in the southern hemisphere. We believe the most likely explanation for this is that, prior to the {\it Gaia\/} era, the available proper motion surveys in the Southern hemisphere have been less extensive \citep[see, e.g.,][for a discussion of this in the context of the TESS Input Catalog]{Stassun:2018}.

Given the similarity of the KELT and TESS pixel scales and the significant overlap of sky coverage (see Figure \ref{fig:kelt_fields}), the KELT FP Catalog provides a pre-vetted set of false positives for TESS. Public knowledge of these data will help to minimize duplication of follow-up observations during the TESS era.

\section{Summary}

The KELT transiting exoplanet discovery process is presented including our follow-up observation process that determines if a KELT-detected candidate event is caused by a transiting planetary mass companion, or if it was caused by an FP or FA. We also describe our large world-wide network of professional, student, and highly capable citizen astronomer photometric and spectroscopic follow-up observation partners. Tools developed out of the KELT project to aid in selecting, scheduling, and reducing follow-up observations, and that keep our KELT-FUN team working efficiently, are also presented.

The KELT-FUN team has been conducting follow-up observations since 2011, which have so far produced more than 20 transiting exoplanet planet discoveries, 1,128 FP confirmations, and $\sim 450$ FAs. The planet discovery rate is $\sim 1.3\%$ after human vetting of the KELT candidate events, and $\sim 2\%$ after FAs have been eliminated. The high FA rate is due to our slight reduction of KELT detection thresholds to minimize the chance of throwing out candidate events that are actually caused by a transiting exoplanet. These more aggressive detection thresholds are possible because of our strong follow-up network. The relative high FP rate is due to the large KELT pixels (and the resulting large $3\arcmin$ photometric apertures) and the relatively low KELT photometric precision, but is again facilitated by size and dedication of the KELT-FUN team.

The FPs have been classified into four spectroscopic and five photometric categories. The NEB and SB1 categories are the dominant photometric and spectroscopic categories, respectively, with NEBs being the category with the most FPs due to our general photometry-first follow-up approach. The giant FP category has only 29 total FPs, indicating that the reduced proper motion cut technique used to minimize the number of spectroscopic giants in the KELT input catalog performs well.

We expect that the KELT survey will continue into the era of TESS for an indeterminate amount of time. There will be regions of the sky not fully covered by TESS during the primary mission in which KELT can continue to confirm new planet discoveries. Furthermore, we expect that continuing to build upon the already long time baseline of KELT data will yield valuable results for transiting exoplanet science as well as other ancillary science. The success of KELT-FUN shows the value of an organized and motivated combination of professional, student, and citizen astronomers, and such efforts will play an important role in confirming TESS Objects of Interest (TOIs) as planets. While the TESS mission is organizing follow-up observers under the TESS Follow-up Observing Program (TFOP), we expect that KELT-FUN will continue into the era of TESS to support the continued KELT survey, and to follow-up TOIs in ways that are complementary to TFOP. One such planned complementary program intends to combine the long time baseline of the KELT data with TESS single transit detections and KELT-FUN observations to confirm long-period giant planets (Yao et al., in prep).

The KELT FP catalog has been published to help minimize duplication of follow-up observation efforts by current and future transiting planet wide-field surveys such as TESS. We encourage other transit surveys to make their catalogs of FPs public to help increase the efficiency of planet confirmation for the TESS mission and other wide-field transiting exoplanet surveys, and for the benefit of the exoplanet community in general.

\acknowledgements
The authors thank the anonymous reviewer and scientific editor for helpful suggestions regarding both form and content.
We also thank other KELT-FUN participants who enabled, gathered and/or reduced data for this project, including Michael Endl, Chas Beichman, Lars Buchhave, Debra Fischer, Ian Crossfield, Rahul Patel, and many others.
This project makes use of data from the KELT survey, including support from The Ohio State University, Vanderbilt University, and Lehigh University, along with the KELT follow-up collaboration.
Work performed by J.E.R. was supported by the Harvard Future Faculty Leaders Postdoctoral fellowship.
D.J.S. and B.S.G. were partially supported by NSF CAREER Grant AST-1056524. Early work on KELT-North was supported by NASA Grant NNG04GO70G.
Work by S.V.Jr. is supported by the National Science Foundation Graduate Research Fellowship under Grant No. DGE-1343012 and the David G. Price Fellowship in Astronomical Instrumentation.
Work by G.Z. is provided by NASA through Hubble Fellowship grant HST-HF2-51402.001-A awarded by the Space Telescope Science Institute, which is operated by the Association of Universities for Research in Astronomy, Inc., for NASA, under contract NAS 5-26555.
K.K.M. acknowledges the support of the Theodore Dunham, Jr. Fund for Astronomical Research and the NASA Massachusetts Space Grant consortium.
J.R.C. acknowledges partial support from NASA grant NNX14AB85G.
D.W.L., K.A.C., and S.N.Q. acknowledge partial support from the TESS Mission.
M.D.J., D.C.S., and E.G.H. thank the Brigham Young University College of Physical and Mathematical Sciences for continued support of the WMO and OPO research facilities.
This paper includes data taken at The McDonald Observatory of The University of Texas at Austin.
Research at the Phillips Academy Observatory is supported by the Israel Family Foundation and the Abbot Academy Association.
This work is partially based on observations obtained with the 1.54-m telescope at Estaci\'{o}n Astrof\'{i}sica de Bosque Alegre dependent on the National University of C\'{o}rdoba, Argentina.
This research made use of Montage. It is funded by the National Science Foundation under Grant Number ACI-1440620, and was previously funded by the National Aeronautics and Space Administration's Earth Science Technology Office, Computation Technologies Project, under Cooperative Agreement Number NCC5-626 between NASA and the California Institute of Technology.
This work makes use of observations from the LCOGT network.
We thank T\"{U}B\.{I}TAK for the partial support in using T100 telescope with project number 16CT100-1096. Authors from Ankara University also acknowledge the support by the research fund of Ankara University (BAP) through the project 13B4240006.
This work has made use of NASA's Astrophysics Data System, the Extrasolar Planet Encyclopedia, the NASA Exoplanet Archive, the SIMBAD database operated at CDS, Strasbourg, France, and the VizieR catalogue access tool, CDS, Strasbourg, France. We make use of Filtergraph, an online data visualization tool developed at Vanderbilt University through the Vanderbilt Initiative in Data-intensive Astrophysics (VIDA).
We also used data products from the Widefield Infrared Survey Explorer, which is a joint project of the University of California, Los Angeles; the Jet Propulsion Laboratory/California Institute of Technology, which is funded by the National Aeronautics and Space Administration; the Two Micron All Sky Survey, which is a joint project of the University of Massachusetts and the Infrared Processing and Analysis Center/California Institute of Technology, funded by the National Aeronautics and Space Administration and the National Science Foundation.
MINERVA is a collaboration among the Harvard-Smithsonian Center for Astrophysics, The Pennsylvania State University, the University of Montana, and the University of New South Wales. MINERVA is made possible by generous contributions from its collaborating institutions and Mt. Cuba Astronomical Foundation, The David \& Lucile Packard Foundation, National Aeronautics and Space Administration (EPSCOR grant NNX13AM97A), The Australian Research Council (LIEF grant LE140100050), and the National Science Foundation (grants 1516242 and 1608203). Any opinions, findings, and conclusions or recommendations expressed are those of the author and do not necessarily reflect the views of the National Science Foundation.
The Digitized Sky Surveys were produced at the Space Telescope Science Institute under U.S. Government grant NAG W-2166. The images of these surveys are based on photographic data obtained using the Oschin Schmidt Telescope on Palomar Mountain and the UK Schmidt Telescope. The plates were processed into the present compressed digital form with the permission of these institutions. The National Geographic Society - Palomar Observatory Sky Atlas (POSS-I) was made by the California Institute of Technology with grants from the National Geographic Society. The Second Palomar Observatory Sky Survey (POSS-II) was made by the California Institute of Technology with funds from the National Science Foundation, the National Geographic Society, the Sloan Foundation, the Samuel Oschin Foundation, and the Eastman Kodak Corporation. The Oschin Schmidt Telescope is operated by the California Institute of Technology and Palomar Observatory. The UK Schmidt Telescope was operated by the Royal Observatory Edinburgh, with funding from the UK Science and Engineering Research Council (later the UK Particle Physics and Astronomy Research Council), until 1988 June, and thereafter by the Anglo-Australian Observatory. The blue plates of the southern Sky Atlas and its Equatorial Extension (together known as the SERC-J), as well as the Equatorial Red (ER), and the Second Epoch [red] Survey (SES) were all taken with the UK Schmidt. All data are subject to the copyright given in the copyright summary\footnote{http://archive.stsci.edu/dss/copyright.html}. Copyright information specific to individual plates is provided in the downloaded FITS headers. Supplemental funding for sky-survey work at the ST ScI is provided by the European Southern Observatory.

\software{AstroImageJ \citep{Collins:2017}, FilterGraph \citep{Burger:2013}, ISIS \citep{Alard:1998, Alard:2000}, TAPIR \citep{Jensen:2013}}

\bibliographystyle{aasjournal}

\bibliography{KELT-FP}

\end{document}